\documentclass[longauth]{aa} 

\usepackage{graphicx}
\usepackage{txfonts}
\usepackage{pifont}
\newcommand{\cmark}{\ding{51}}%
\newcommand{\xmark}{\ding{55}}%
\usepackage[colorlinks=true, allcolors=blue]{hyperref}
\usepackage{placeins}
\usepackage[rightcaption]{sidecap}
\usepackage[english]{babel}
\usepackage{array}
\usepackage{capt-of}
\usepackage[normalem]{ulem}
\usepackage{lscape}
\usepackage{multirow}
\usepackage{float}
\usepackage{placeins}

\begin{document} 

   \title{JOYS: MIRI/MRS spectroscopy of gas-phase molecules from the high-mass star-forming region IRAS 23385+6053}
   \titlerunning{MIRI/MRS molecular spectroscopy of IRAS 23385+6053}

   \author{
L.~Francis\inst{1} \and
M.~L.~van~Gelder\inst{1} \and
E.~F.~van~Dishoeck\inst{1, 3} \and
C.~Gieser\inst{3, 2} \and
H.~Beuther\inst{2} \and
L.~Tychoniec\inst{4,1} \and
G.~Perotti\inst{2} \and
A.~Caratti~o~Garatti\inst{6} \and
P.~J. Kavanagh\inst{5} \and
T.~Ray\inst{7} \and
P.~Klaassen\inst{8} \and
K.~Justtanont\inst{9} \and
H.~Linnartz\inst{10} \and
W.~R. M. Rocha\inst{1,10} \and
K.~Slavicinska\inst{1, 10} \and
M.~Güdel\inst{11, 2, 12} \and
T.~Henning\inst{2} \and
P.-O.~Lagage\inst{13} \and
G.~Östlin\inst{14}
          }     

\institute{
Leiden Observatory, Leiden University, PO Box 9513, 2300 RA Leiden, The Netherlands \\ email{francis@strw.leidenuniv.nl} \and
Max Planck Institute for Extraterrestrial Physics, Gießenbachstraße 1, 85749 Garching bei München \and
Max Planck Institute for Astronomy, Königstuhl 17, 69117 Heidelberg, Germany \and
European Southern Observatory, Karl-Schwarzschild-Strasse 2, 85748 Garching bei München, Germany \and
Department of Experimental Physics, Maynooth University-National University of Ireland Maynooth, Maynooth, Co Kildare, Ireland \and
INAF-Osservatorio Astronomico di Capodimonte, Salita Moiariello 16, I-80131 Napoli, Italy \and
School of Cosmic Physics, Dublin Institute for Advanced Studies, 31 Fitzwilliam Place, Dublin 2, Ireland \and
UK Astronomy Technology Centre, Royal Observatory Edinburgh, Blackford Hill, Edinburgh EH9 3HJ, UK \and
Department of Space, Earth and Environment, Chalmers University of Technology, Onsala Space Observatory, 439 92 Onsala, Sweden \and
Laboratory for Astrophysics, Leiden Observatory, Leiden University, PO Box 9513, NL 2300 RA Leiden, The Netherlands \and
Department of Astrophysics, University of Vienna, Türkenschanzstr. 17, 1180 Vienna, Austria \and
ETH Zürich, Institute for Particle Physics and Astrophysics, Wolfgang-Pauli-Str. 27, 8093 Zürich, Switzerland \and
Université Paris-Saclay, Université de Paris, CEA, CNRS, AIM, 91191 Gif-sur-Yvette, France \and
Department of Astronomy, Oskar Klein Centre, Stockholm University, 106 91 Stockholm, Sweden
}

\date{Received 29 September 2023; accepted January 9 2024}

 
  \abstract
   {Space-based mid-infrared (IR) spectroscopy is a powerful tool for the characterization of important star formation tracers of warm gas which are unobservable from the ground. The previous mid-IR spectra of bright high-mass protostars with the Infrared Space Observatory (ISO) in the hot-core phase typically show strong absorption features from molecules such as CO$_2$, C$_2$H$_2$, and HCN. However, little is known about their fainter counterparts at earlier stages.}  
   {We aim to characterize the gas-phase molecular features in {\it James Webb} Space Telescope Mid-Infrared Instrument Medium Resolution Spectrometer (MIRI/MRS) spectra of the young and clustered high-mass star-forming region IRAS 23385+6053.}
   {Spectra were extracted from several locations in the MIRI/MRS field of view, targeting two mid-IR sources tracing embedded massive protostars as well as three H$_2$ bright outflow knots at distances of $>8000$ au from the multiple. Molecular features in the spectra were fit with local thermodynamic equilibrium (LTE) slab models, with their caveats discussed in detail.}
   {Rich molecular spectra with emission from CO, H$_2$, HD, H$_2$O, C$_2$H$_2$, HCN, CO$_2$, and OH are detected towards the two mid-IR sources. However, only CO and OH are seen towards the brightest H$_2$ knot positions, suggesting that the majority of the observed species are associated with disks or hot core regions rather than outflows or shocks. The LTE model fits to $^{12}$CO$_{2}$, C$_{2}$H$_{2}$, HCN emission suggest warm $120-200$ K emission arising from a disk surface around one or both protostars. The abundances of CO$_2$ and C$_2$H$_2$ of $\sim 10^{-7}$ are consistent with previous observations of high-mass protostars. Weak $\sim500$ K H$_2$O emission at $\sim$ 6-7 $\mu$m is detected towards one mid-IR source, whereas $250-1050$ K H$_2$O absorption is found in the other. The H$_2$O absorption may occur in the disk atmosphere due to strong accretion-heating of the midplane, or in a disk wind viewed at an ideal angle for absorption. CO emission may originate in the hot inner disk or outflow shocks, but NIRSpec data covering the 4.6 $\mu$m band head are required to determine the physical conditions of the CO gas, as the high temperatures seen in the MIRI data may be due to optical depth. OH emission is detected towards both mid-IR source positions and one of the shocks, and is likely excited by water photodissociation or chemical formation pumping in a highly non-LTE manner.} 
   {The observed molecular spectra are consistent with disks having already formed around two protostars in the young IRAS 23385+6054 system. Molecular features mostly appear in emission from a variety of species, in contrast to the more evolved hot core phase protostars which typically show only absorption; however, further observations of young high-mass protostars are needed to disentangle geometry and viewing angle effects from evolution.}

\maketitle
%

\section{Introduction}

Gas-phase molecular emission and absorption is ubiquitous in the spectra of the disks and outflows associated with young protostars. Analysis of their spectra provides a valuable probe into the physical conditions where molecules are formed, and the mechanism for their excitation. Mid-infrared (IR) spectroscopy, in particular, offers a unique window for observing rovibrational molecular features \citep[e.g.][]{vanDishoeck2004,vanDishoeck2023,Hollenbach1989}. Key star formation tracers such as C$_2$H$_2$, CO$_2$, and CH$_4$ are among the most abundant carbon and oxygen carrying species, but they do not have observable rotational spectra due to their lack of a permanent electric dipole moment, and thus they can only be observed in the mid-IR through their rovibrational transitions. Furthermore, observations from space telescopes are essential for observing molecules such as CO$_2$ and H$_2$O, whose presence in our atmosphere makes their detection from the ground difficult.

With the launch of the {\it James Webb} Space Telescope (JWST), space-based mid-IR spectroscopy is once again possible with the Mid-Infrared Instrument (MIRI; \citep{Rieke2015,Wright2015,Wright2023}  Medium Resolution Spectrometer (MRS; \citealt{Wells2015}). JWST provides several orders of magnitude in improvements to sensitivity and spatial resolution over previous observatories such as {\it Spitzer} and the Infrared Space Observatory (ISO). The improved spectral resolution of MIRI/MRS of $R=\Delta \lambda / \lambda \sim3400-1600$ \citep{Argyriou2023a} versus 50-600 for {\it Spitzer} significantly boosts the line to continuum ratio for gas-phase lines, although MIRI/MRS does not reach as high a resolution as the $R\sim 30000$ of ISO-SWS. The unmatched sensitivity of JWST also enables spectroscopy of far fainter and more distant targets than possible before. This is particularly useful for the study of massive protostars, which are rarer than their low-mass siblings and have short evolutionary timescales of a few 100,000 yr \citep[e.g.][]{Mottram2011,Motte2018}. Furthermore, the sub-arcsecond spatial resolution of JWST is valuable for densely clustered star-forming regions where high-mass stars preferentially form. 

Past mid-IR studies of high-mass protostars have largely focussed on those in the more evolved `hot core' phase, where significant heating of the envelope by a massive protostar produces diverse complex organic molecules (COMs: carbon-bearing molecules with six or more atoms) and the formation of compact HII regions \citep{Gerner2014,Cesaroni1997}. Prior to the hot core phase, the gas-phase COM abundance is lower. Towards these protostars, 
transitions of molecules such as CO$_2$, C$_2$H$_2$, HCN, and H$_2$O are typically seen in absorption \citep[e.g.][]{Evans1991,Helmich1996a,Boonman2003a,Boonman2003b,Boonman2003c}. High-resolution spectroscopy at up to $R=100000$ with the Stratospheric Observatory for Infrared Astronomy Echelon-Cross-Echelle Spectrograph (SOFIA/EXES) and Very Large Telescope Cryogenic High-resolution Infrared Echelle Spectrograph (VLT/CRIRES) suggests that the absorption features often arise from the surface layers against the accretion-heated midplane of a massive protostellar disk \citep{Indriolo2020,Barr2020,Barr2022}. However, mid-IR molecular emission has also occasionally been detected in high-mass protostars, where it is associated with warm gas in the outflow excited by mid-IR pumping (e.g. the outflows of Cepheus A, \citep{Sonnentrucker2006,Sonnentrucker2007} and Orion-IRc2/KL, \cite{Boonman2003c}). 

Mid-IR molecular emission from embedded disks of both low- and high-mass protostars has rarely been detected, likely due to extinction from the envelope and lower sensitivities of past observatories \citep{Lahuis2010}. Recent MIRI/MRS observations of class 0 protostars have identified gas-phase features of CO and H$_2$O possibly associated with the disk \citep{Yang2022}, and a stronger line forest from a variety of species in the eruptive protostar EX Lup \citep{Kospal2023}. In the more evolved class II stage, molecular emission is much more frequently detected and better characterized, and previous ground-based and {\it Spitzer} surveys have shown it to originate from the hot inner disk and warm disk surface \citep[e.g.][]{Blake2004,Carr2008,Carr2011,Salyk2011,Pontoppidan2010a,Pontoppidan2014,Banzatti2022,Banzatti2023}. JWST observations of class II disks have confirmed this, and also suggested links between the inner disk molecular inventory and dust transport within the disk \citep{Grant2023,Tabone2023,Gasman2023,Perotti2023,Banzatti2023}.

In this paper, we present an analysis of gas phase molecular features in JWST/MIRI observations of the high mass star-forming region IRAS 23385+6053. This paper is part of a series on the first results from the JWST Observations of Young Protostars (JOYS) collaboration: an overview of the IRAS23385+6053 MIRI/MRS observations and the detection of the accretion-tracing Humpreys $\alpha$ line has been presented by \citet{Beuther2023}, an analysis of the outflows and warm gas traced in the mid-IR and sub-mm  by \citet{Gieser2023}, and the ice absorption features by \citet{Rocha2023}. Some of the data presented in this paper have also been discussed by \cite{vanDishoeck2023} in the context of the astrochemistry of planet-forming disks.   

IRAS 23385+6053 is located in the outer Galaxy at a distance of 4.9 kpc \citep{Molinari1998} and a Galactocentric distance of 11 kpc, with a luminosity of $\sim 3000~ L_\odot$ and total envelope mass of 510 $M_\odot$ \citep{Cesaroni2019}. Previous observations suggest that IRAS 23385+6053 is extremely young, with a low bolometric temperature of $\sim40$ K \citep{Fontani2004}, and no observed HII regions or free-free emission. Furthermore, IRAS 23385+6053 is not as rich in COMs as typical hot core regions \citep{Cesaroni2019}, consistent with a young age. The star formation in IRAS 23385+6053 is highly clustered, with up to 6 different cores detected in the sub-mm \citep{Cesaroni2019} and a complex system of at least three high velocity outflows \citep{Beuther2023}. The detection of large-scale gas motions suggests a large disk may be present around a $\sim 9~M_\odot$ star within one of the brightest cores \citep{Cesaroni2019}.

The remainder of this paper is organized as follows: in Section \ref{sec:obs_methods}, we discuss the data reduction, extraction of spectra in apertures containing the various protostars and outflow knots, and the LTE slab models used to identify and fit the observed molecular features. In Section \ref{sec:results}, we then list the identified species and the results of the LTE model fits. We continue with a discussion of the origin of each detected species and the components of the protostellar system they most likely trace in Section \ref{sec:disc}. Finally, we summarize our overall model for the molecular features in IRAS 23385+6053, and conclude with some remarks on the future outlook for JWST studies of high mass star formation in Section \ref{sec:summary}.

\section{Observations and methods}
\label{sec:obs_methods}

\subsection{Data reduction}

The full details of the data reduction are described in \cite{Beuther2023}. In summary, we have run the JWST calibration pipeline \citep{Bushouse2022} using the reference context {\tt jwst$\_$1017.pmap} of the JWST Calibration Reference Data System \citep[CRDS;][]{Greenfield2016}. Additionally, we have applied an astrometric correction using GAIA DR3 stars in the simultaneous MIRI imaging field of view. A dedicated background field was taken off-source. However, due to significant astronomical emission in the dedicated background observation, subtracting this background from the science data resulted in negative fluxes. The telescope background was therefore estimated by extracting a spectrum from within the primary science field of view, but off-source from the main infrared continuum sources at the position within the IFU where the background flux was the lowest (R.A. (J2000) 23$^{\rm h}$40$^{\rm m}$54.15$^{\rm s}$, Dec (J2000) 61$^{\rm d}$10$^{\rm m}$26.96$^{\rm s}$). The aperture size was the same as was used for extracting the science data (Section \ref{ssec:overview_and_extraction}). 
The background subtraction also resulted in the subtraction of the strong PAH emission features around 8.6~$\mu$m and 11.3~$\mu$m since their emission was about equally strong in the offset position, although some residuals remain. However, the molecular emission discussed in this paper does not overlap with any of the PAH bands and is therefore not affected by them.

\subsection{Source overview and spectra extraction}
\label{ssec:overview_and_extraction}
\begin{figure*}[htb]
    \centering
    \includegraphics[width=\textwidth]{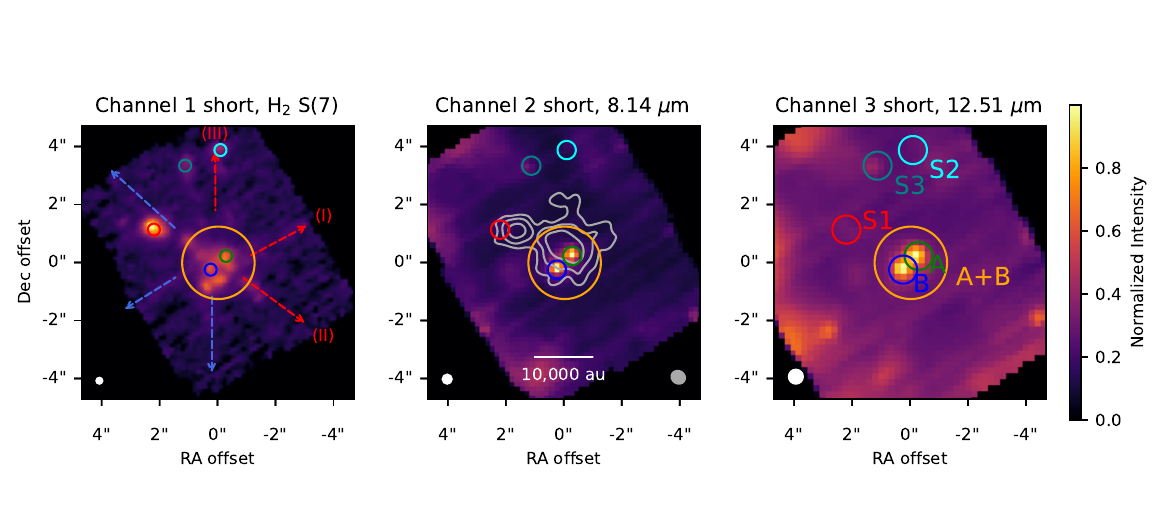}
    \caption{First three channels of MIRI/MRS observations of IRAS 23385+6053. The left panel shows the H$_2$ S(7) transition in channel 1 at 5.511 $\mu$m, while the centre and right panels show the averaged continuum in channels 2 and 3. The colour scale in all panels is normalized to the peak intensity. The arrows in the left panel indicate the directions and velocity shift of the major outflows proposed by \cite{Beuther2023}. The extraction apertures for our spectra are shown in each panel by the coloured circles. Aperture A+B has a constant radius of 2.5\arcsec, while all other apertures have a radius that scales with wavelength as $r=2(1.22\lambda/6.5\mathrm{m})$. The grey contours overlaid on channel 2 are 1.3 mm continuum emission \citep{Gieser2021}. The white circle in the bottom left indicates the mean beam size of JWST in the given channel, while the grey ellipse in the second panel indicates the FWHM of the mm beam.}
    \label{fig:apertures}
\end{figure*}

An overview of our MIRI/MRS observations of IRAS 23385+6053 has been presented by \cite{Beuther2023}, which we briefly summarize here.  In Figure \ref{fig:apertures}, the emission from the H$_2$ S(7) line in channel 1 is shown in the first panel, while the continuum from the second and third MIRI/MRS channels is shown by the colour scale. For reference, we indicate the putative directions of the major outflows in the left panel. At mid-IR wavelengths, two mid-IR sources with a separation of $\sim0.67\arcsec$ / $3280$ au are detected; it is not known if these sources are gravitationally bound. Following the naming of \cite{Beuther2023}, we refer to the north-west and south-east sources as A and B respectively. source A and B are clearly resolved in channels 1 and 2 (4.9-11.7 $\mu$m) but begin to become blended in channel 3 (11.55–17.98 $\mu$m) and are completely blended in channel 4 (17.7–27.9 $\mu$m). Spatially extended H$_2$ emission tracing the outflows is detected in the S(1) to S(8) transitions over the majority of the MIRI field of view \citep{Beuther2023}. On the basis of their large projected separation and the presence of ice and silicate absorption features towards both sources, \cite{Beuther2023} argue that source A and B are two distinct protostars, which are likely responsible for driving at least two of the observed outflows in this region. 

IRAS 23385+6053 has also been observed at mm wavelengths as part of the CORE programme \citep{Beuther2018,Cesaroni2019,Gieser2021}, where extended continuum emission likely tracing the positions of several deeply embedded protostars is detected (grey contours, middle panel of Figure \ref{fig:apertures}). source A is seen to be closely aligned with an existing mm peak, but source B is clearly offset, although still located within the cold envelope. Furthermore, although mid-IR continuum emission is only detected from source A and B, up to 6 cores have been identified from sub-mm molecular emission by \cite{Cesaroni2019}. The lack of significant mid-IR emission towards the other cores may reflect a difference in mass or evolutionary state of protostars in the cluster, or simply a lack of sensitivity, as suggested by \cite{Beuther2023}.

As mid-IR molecular features have previously been detected at both the locations of protostars and in the outflows \citep{Beuther2023,Gieser2023}, we extract spectra towards sources A and B, as well as the three brightest H$_2$ knots (see Figure \ref{fig:apertures}), which are named S1, S2, and S3 following \cite{Gieser2023}. We note that the brightest knot (S1) 
is close to a bright mm peak possibly tracing a prestellar core or embedded protostar, and that the H$_2$ emission here is seen over an extensive velocity range of $\pm 140$ km s$^{-1}$, suggesting strong outflow activity \citep{Beuther2023}. Our extraction apertures for each source are shown by the circles in Figure \ref{fig:apertures}. Since the spatial resolution of MIRI/MRS varies significantly from $\sim 0.27$\arcsec~in channel 1 to $\sim 1.0$\arcsec~in channel 4 \citep{Law2023}, we sum the flux in a `conical'  aperture whose radius scales with wavelength as $r=1.22 (\lambda/6.5\mathrm{m})$. Because of the blending of sources A and B in channels 3 and 4, we also extract a spectrum in a large aperture containing both sources with a fixed diameter of 2.5\arcsec, identical to that used by \cite{Beuther2023}. To all of our spectra, we apply a residual fringe correction procedure to remove fringing behaviour not corrected for by the pipeline \citep{KavanaghinPrep}.

\subsection{Baseline subtraction and ice absorption correction}
\label{ssec:baseline}

To characterize the molecular features, we first perform a baseline subtraction using univariate spline fits to line-free regions, in order to remove the thermal continuum, PAH emission, and ice absorption. We also model the thermal continuum using a second order polynomial fit to areas assumed to be free of ice-absorption. Two examples of local baselines and thermal continuum fits to the $\sim6.5$ $\mu$m and $\sim15$ $\mu$m regions are shown in Figure \ref{fig:cont_spline} of the appendix. More global continuum fits to study ices are presented in \cite{Rocha2023}. 

\subsection{LTE slab models}
\label{ssec:slab_models}

We fitted the continuum-subtracted molecular emission using the local thermodynamic equilibrium (LTE) slab models described in \cite{Grant2023,Tabone2023}, which are similar in functionality to the existing \texttt{slabspec} code \citep{Salyk2020}. The molecular emission is assumed to originate in a plane-parallel uniform cylindrical slab of radius $R$ and column density $N$, with a gas at a single temperature $T_\mathrm{ex}$. In reality, protostellar environments should contain gas over a wide range of densities and temperatures in the disk, outflow, and envelope which such simple models cannot capture, and non-LTE effects may be particularly important for regions with low density and strong radiative pumping (e.g. outflow shocks). In particular, the effect of mid-IR pumping in particular can result in model excitation temperatures and column densities which do not reflect the kinetic temperature and column density of that gas, which we discuss further in Section \ref{ssec:origins_disc}. Nevertheless, LTE models are still valuable for identifying and characterizing the molecular emission; we discuss caveats of their use in Section \ref{sec:disc}.

The molecular spectroscopy data for the models (transition wavelengths, Einstein $A$ coefficients, statistical weights, and partition functions) are taken from the \texttt{HITRAN} database \citep{Gordon2022}. This information is used in combination with an assumed $T_\mathrm{ex}$ and $N$ to solve for the level populations. The opacity at line centre is determined assuming a turbulent broadening by a Gaussian with full width at half maximum of $\Delta V=4.7$ km s$^{-1}$, which has typically been assumed for LTE models of molecular emission in T-Tauri disks \citep[e.g.][]{Salyk2011}. The typical mm line widths in the CORE survey are between 2-5 km s$^{-1}$, so this assumption is reasonable for IRAS 23385+6035 \citep[See Figure 7 of][]{Cesaroni2019}. The wavelength dependent line optical depth is computed on a high resolution $(\lambda/\Delta \lambda)=10^6$ grid assuming a Gaussian line profile, as there may be significant overlap between closely spaced transitions from molecules with a $Q$-branch \citep{Tabone2023}. The model flux $F(\lambda)$ is then computed as

\begin{equation}
\label{eqn:model_flux}
    F(\lambda) = \pi \left(\frac{R}{d}\right)^2 B_\nu (T) (1 - e^{-\tau (\lambda)}),
\end{equation}

where $R$ is the radius of the emitting slab, $d=4.9$ kpc is the distance to the source, $B_\nu$ is the Planck function, and $\tau(\lambda)$ is the line optical depth. The model is then convolved to the average MIRI instrumental resolution in each sub-band \citep{Labiano2021,Jones2023}. For comparisons with the observations, we velocity shift the convolved synthetic spectra by $v_\mathrm{LSR}=-50.2$ km s$^{-1}$ for sources A and B \citep{Beuther2018} and a by-eye-estimated -100 km s$^{-1}$ for source S1, then finally resample at the observed wavelengths of the MIRI spectra. 

Correction for extinction from the local molecular cloud and protostellar envelope is important to obtain accurate column densities from our molecular emission models. We follow the approach of \cite{Gieser2023}, who corrected the IRAS 23385+6053 spectra using the extinction curves of \cite{McClure2009} with a $K$-band extinction of $A_K=7.0$. However, we make one important modification to more accurately for the effect of ice absorption. The extinction curves of \cite{McClure2009} include features from cloud absorption of H$_2$O and CO$_2$ ice, which may vary significantly from cloud to cloud in shape and depth, and extincts the molecular emission with a strong wavelength dependence. We thus remove the ice features from these extinction curves using a local polynomial fit to ice-free regions. To then include the effect of extinction by ices unique to IRAS 23385+6053 (located in either the protostellar envelope or surrounding molecular cloud),
we calculate the optical depth of the ice feature $\tau=\exp{(-F_\mathrm{cont}/F_\mathrm{baseline}})$ as a function of wavelength, and apply it as an additional correction factor to our molecular emission models. An example of this correction is shown for CO$_2$ in Figure \ref{fig:CO2_ice_corr}, where the ice absorption is seen to significantly weaken the $Q$ and $P$-branch emission at $\lambda > 14.9~\mu$m. Similarly, from 5.5 to 7.8 $\mu$m, there is also strong absorption from H$_2$O, NH$_4^+$, and CH$_4$ ice (lower panel of Figure \ref{fig:cont_spline}) which overlaps with the H$_2$O and CH$_4$ emission features, and thus an analogous correction is applied to the models. An important caveat of this approach is the implicit assumption that all of the absorbing ice lies in front of all of the emitting gas. This assumption may not be completely correct if a large fraction of the emitting gas is in an outflow outside the protostellar envelope, and may be further complicated by the presence of multiple sources experiencing varying degrees of extinction within a single aperture in the case of IRAS 23385+6053.

Molecular absorption features are detected in one case (H$_2$O in source B), and modelled in a similar way using LTE slab models based on the work of \citep{Helmich1996b}. We follow the same procedure outlined above, but only calculating the line optical depth as a function of wavelength, followed immediately by convolution to the MIRI resolution, velocity shifting, and re-sampling. The models are compared to the observed optical depth, which is calculated as $\tau_\mathrm{obs} = -\log(F_\mathrm{obs}/F_\mathrm{baseline})$. Unlike the case of emission, corrections for extinction by the cloud and envelope ices are not applied, as these do not influence the observed optical depth of the transitions. 

\subsection{Model-fitting procedure}

To identify best-fit emission models, for each molecule detected in source A, B, and S1 (see Table \ref{tab:detected_molecules}), we compute a grid of models with varying $N$ and $T_\mathrm{ex}$, the details of which are provided in Appendix \ref{sec:app_chi2_maps}. For each model, we calculate
\begin{equation}
    \chi^2=\sum_{i=1}^{N_\mathrm{obs}}((F_{\mathrm{obs},i} -F_{\mathrm{model},i})/\sigma)^2,
\end{equation}
where $F_{\mathrm{model},i}$ is determined from equation \ref{eqn:model_flux}, $\sigma$ is estimated from the emission in nearby line-free regions, and a minimization routine is used to determine a value of $R$ that minimizes $\chi^2$. The value of $\sigma$ determined empirically from line-free spectral regions is 0.1 mJy. We only fit regions of the spectrum containing obvious molecular features, thus avoiding contaminating emission from bright atomic lines or other molecules. In some cases, emission from several molecules is blended together (e.g. HCN, OH, C$_2$H$_2$ and CO$_2$, see \cite{Grant2023}). Here, we iteratively fit each molecule in order of decreasing flux, using the residuals of the previous best-fit as the input to the following one. 

A similar approach is used for identifying best-fit absorption models, but in units of model and observed optical depth, with $\chi^2$ calculated as 
\begin{equation}
\chi^2=\sum_{i=1}^{N_\mathrm{obs}}((\tau_{\mathrm{obs},i} - \tau_{\mathrm{model},i})/\tau_\sigma)^2,
\end{equation}
where $\tau_\sigma=\sigma/F_{\mathrm{obs},i}$.
We note that in the case of absorption, there is no emitting area radius $R$ to constrain for the `pencil-beam' of the absorbing column.

Uncertainties on the best-fit $N$ and $T_\mathrm{ex}$ are estimated following \cite{Carr2011} and \cite{Salyk2011} by producing $\chi^2$ maps for the best-fit column density and temperature with contours of the best $R$ value overlaid. Confidence intervals are computed assuming $K=2$ degrees of freedom for the column density and temperature following the approach in \cite{Avni1976}. Specifically, for the map of $\chi^2_\mathrm{red} = \chi^2/K$, we overlay contours relative to the best-fit minimum at levels $\Delta \chi^2_\mathrm{red} = \chi^2_\mathrm{red} - \chi^2_\mathrm{red,min}$. The contour levels corresponding to 1, 2, and 3$\sigma$ are thus $\Delta \chi^2_\mathrm{red} \approx$ 2.3, 6.2, and 11.8.

\section{Results}
\label{sec:results}
\subsection{Detections}
\label{ssec:detections}

\begin{figure*}[htb]
    \centering
    \includegraphics[width=\textwidth]{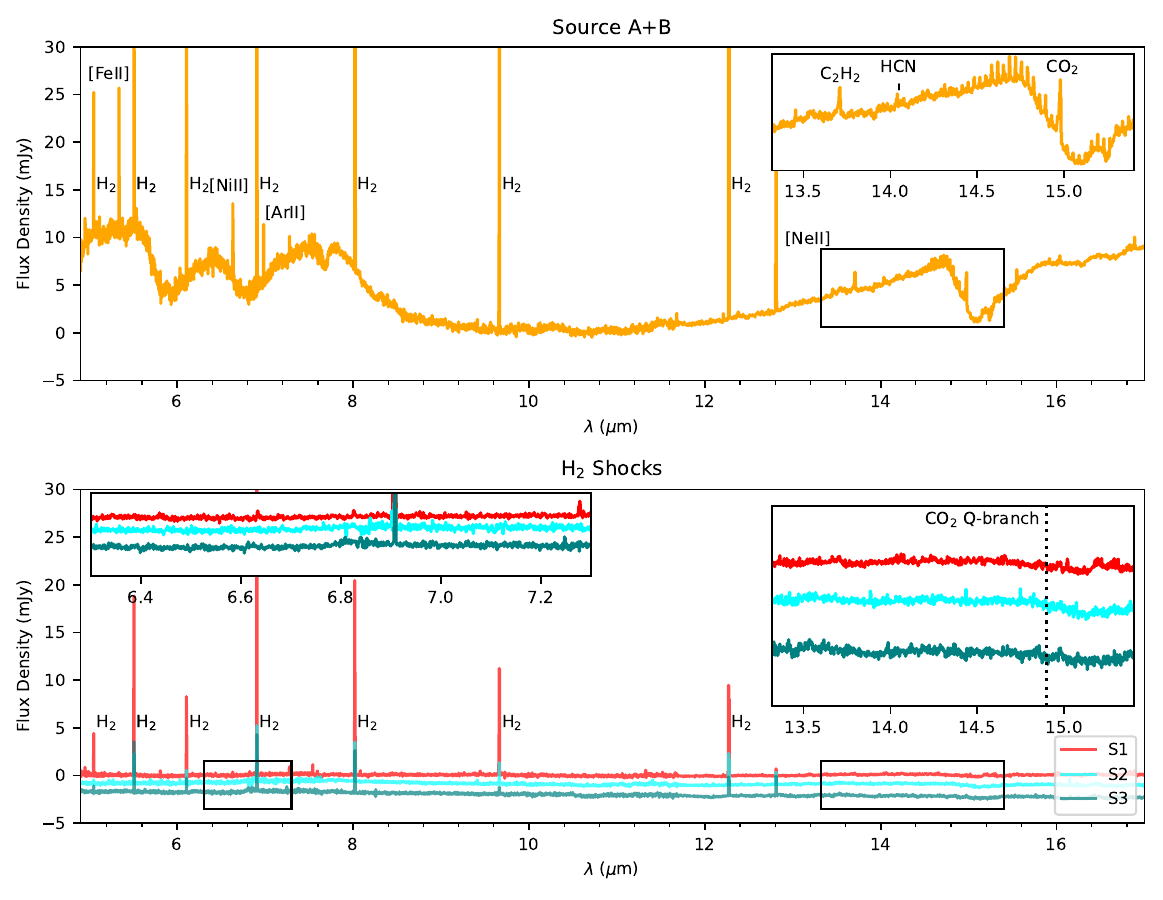}
    \caption{Spectra of sources A+B (top panel) and the brightest H$_2$ knots (lower panel) extracted from the apertures shown in Figure \ref{fig:apertures}. Flux offsets of -2.5 and -5.0 mJy are applied to the spectra of S2 and S3 respectively. In contrast with the A+B spectrum, no molecular lines other than H$_2$, CO, and OH in S1 are detected towards the H$_2$ knots.}
    \label{fig:spectra_overview}
\end{figure*}

An overview of the spectra in the first three MIRI channels is shown in Figure \ref{fig:spectra_overview} for the combined A+B aperture (upper panel) and the H$_2$ knots (lower panel). Mid-IR continuum emission is detected towards A + B at a level of up to $\sim10$ mJy, orders of magnitude lower than the brighter and more evolved high mass sources studied with ISO \citep[e.g.][]{Boonman2003a}, which cannot be observed by JWST due to rapid saturation of the detectors. No continuum emission is seen at the shock positions S1 to S3. Overlaid on the continuum from A + B are strong absorption features from various silicates and ices \citep{Rocha2023}. The aforementioned H$_2$ S(1) to S(8) transitions as well several shock-tracing ionized atomic lines are detected towards all sources (see \citet{Beuther2023}).

\begin{figure*}[htb]
    \includegraphics[width=\textwidth]{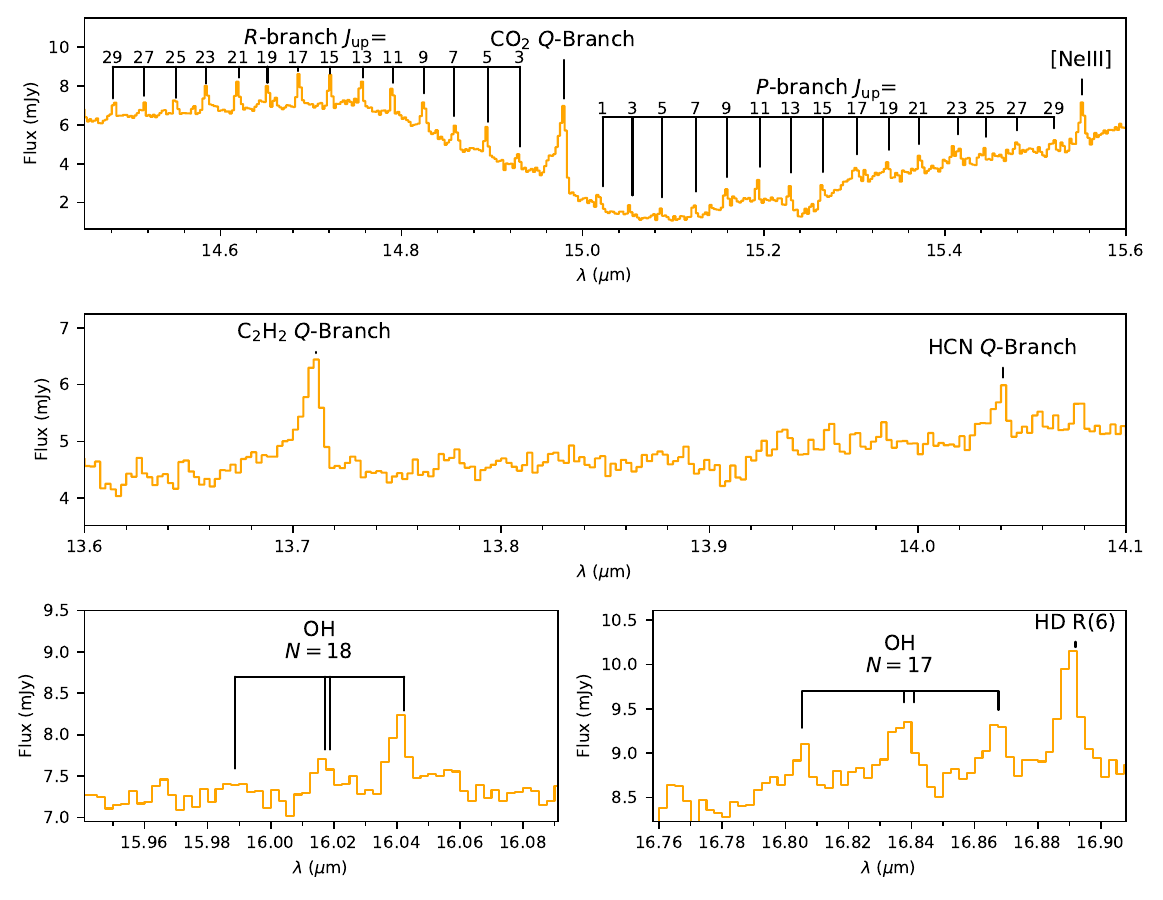}
    \caption{Zoom in on the molecular emission from CO$_2$, C$_2$H$_2$, HCN, and OH emission seen in the A+B aperture (see Figure \ref{fig:apertures}). No continuum subtraction has been applied.}
    \label{fig:AB_CO2_C2H2_HCN}
\end{figure*}

\begin{figure*}[htb]
    \includegraphics[width=\textwidth]{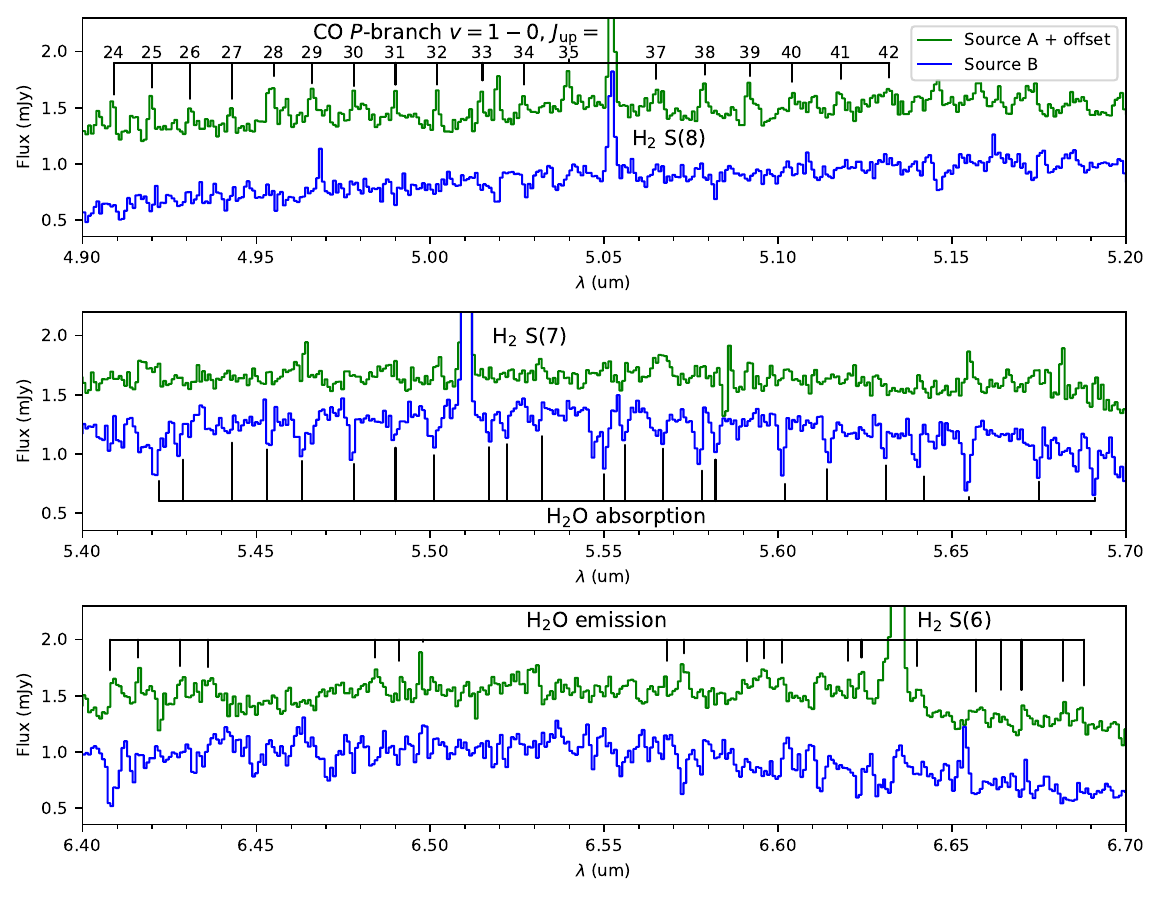}
    \caption{As Figure \ref{fig:AB_CO2_C2H2_HCN}, but for the CO emission and 4.9-6.7 $\mu$m H$_2$O absorption in the spectra of source A (green) and B (blue).}
    \label{fig:A_vs_B_CO_H2O}
\end{figure*}

\begin{figure*}[htb]
    \includegraphics[width=\textwidth]{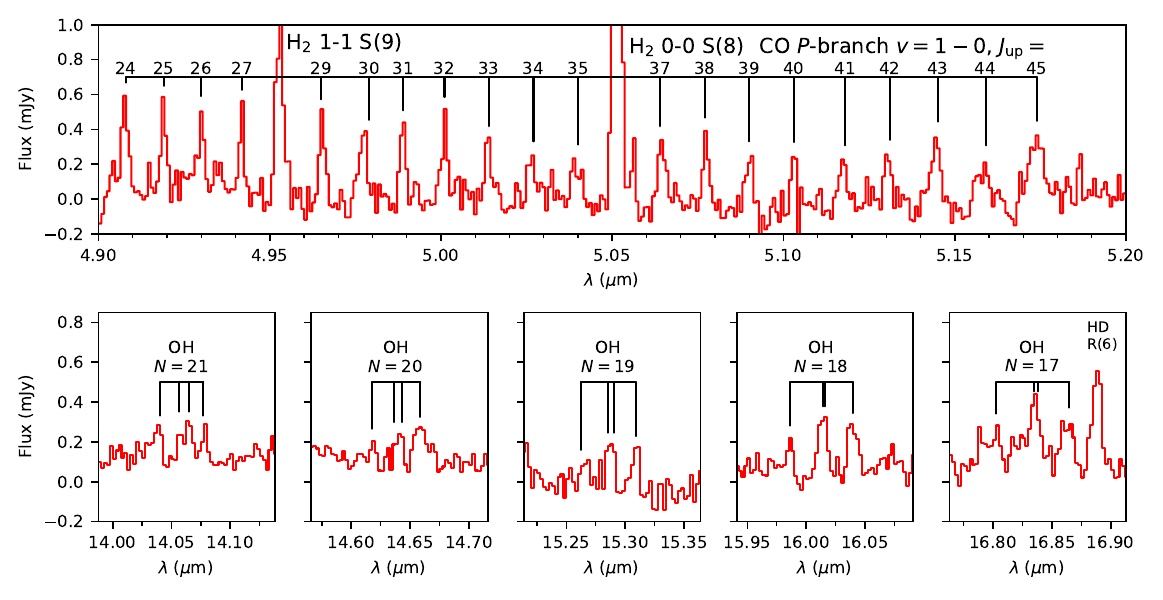}
    \caption{As Figure \ref{fig:AB_CO2_C2H2_HCN}, but for the CO and OH emission spectra in the S1 knot.}
    \label{fig:N_CO_OH}
\end{figure*}

A variety of molecular species are detected towards the different sources, which we summarize in Table \ref{tab:detected_molecules}. As previously noted, H$_2$ is detected over the entire MIRI/MRS field of view, and is clearly associated with the outflows (see \cite{Gieser2023}). The (0-0) R(6), R(5), and R(4) transitions of HD are detected, which in combination with the H$_2$ lines, will be used as a probe of the D/H abundance in a future publication \citep{FrancisinPrep}. Strong molecular emission from CO$_2$, C$_2$H$_2$, and HCN is detected towards source A + B (inset, top panel of Figure \ref{fig:spectra_overview}, and Figure \ref{fig:AB_CO2_C2H2_HCN}), as well as fainter $\sim6~\mu$m and $\sim16~\mu$m H$_2$O features in the individual sources (Figure \ref{fig:A_vs_B_CO_H2O}). Towards the H$_2$ knots no such features are present (insets, lower panel Figure \ref{fig:spectra_overview}), however, emission from CO and OH is detected in source S1 (see also Figure \ref{fig:N_CO_OH}).

\begin{table}
\centering
\caption{Detected molecules.}
\label{tab:detected_molecules}
\begin{tabular}{cccccc}
\hline \hline
Species           &source A & source B  & S1     & S2     & S3 \\
\hline
$^{12}$CO$_{2}$   & \cmark  & \cmark    & \xmark & \xmark & \xmark \\
$^{13}$CO$_{2}$   & \cmark?  & \xmark    & \xmark & \xmark & \xmark \\
C$_{2}$H$_{2}$    & \cmark  & \cmark    & \xmark & \xmark & \xmark \\
$^{13}$CCH$_{2}$  & \cmark?  & \xmark    & \xmark & \xmark & \xmark \\
HCN               & \cmark  & \cmark    & \xmark & \xmark & \xmark \\
H$_{2}$O          & \cmark  & \cmark$^*$& \xmark & \xmark & \xmark \\
H$_2$             & \cmark  & \cmark    & \cmark & \cmark & \cmark \\
HD                & \cmark  & \cmark    & \cmark & \cmark & \cmark \\
CO                & \cmark  & \xmark    & \cmark & \xmark & \xmark \\
OH                & \cmark  & \cmark    & \cmark & \xmark & \xmark \\
\hline
\end{tabular}
\\All species are seen in emission, except where a $^*$ denotes absorption. A `?' denotes a tentative detection.
\end{table}

The spectra towards source A and B are particularly rich in comparison to the H$_2$ knot positions. In Figure \ref{fig:AB_CO2_C2H2_HCN}, we show a zoom-in on the A+B spectra in the 13.6-15.6 $\mu$m range where emission from the CO$_2$, C$_2$H$_2$, and HCN is detected. The CO$_2$ emission is co-located with a strong absorption feature of CO$_2$ ice which extincts the $Q$- and $P$-branches of the emission (Section \ref{ssec:slab_models}). In both sources, weak emission from OH is detected from 14-17 $\mu$m, but most transitions except those at $\sim 16.04$ and $\sim16.84$ $\mu$m are blended with CO$_2$ and HCN emission. The isotopologues $^{13}$CO$_2$ and $^{13}$CCH$_2$ are also detected, but are blended with emission from the $P$-branch of CO$_2$ and the $Q$-branch of C$_2$H$_2$.

Several differences are apparent between the spectra of source A and B as well. In channels 1 and 2 (4.9 to 11.7 $\mu$m), emission from sources A and B is spatially resolved, but partial blending and overlap of our extraction apertures begins to occur in channel 3 (11.55-17.98 $\mu$m). We therefore analyse source A and B independently in channels 1 and 2 only. In source A, rovibrational emission from the CO $P$-branch is detected from 4.9 to 5.2 $\mu$m with $v=1-0,J\rightarrow J+1$ levels up to $J=45$ ($E_\mathrm{up}=9481$ K), indicating extremely energetic excitation conditions. CO emission is entirely absent in source B, however (see Figure \ref{fig:A_vs_B_CO_H2O}). 

H$_2$O is detected in both sources from 4.9 to 8 $\mu$m in its $v_2 (1-0)$ vibrational bending mode, but in emission in source A and in absorption in source B. In source A, a handful of `cold' transitions of pure rotational water emission at $\sim16-18$ $\mu$m are also detected, though blending with source B makes this association ambiguous. 

In the H$_2$ hotspots, fewer molecules are detected. Emission from CO and OH is only seen towards the brightest source S1 (see Figure \ref{fig:N_CO_OH}), likely tracing the presence of strong, dense shocks.

\subsection{LTE slab model fits}
\label{ssec:lte_slab_fits}

As the emission from CO$_2$ C$_2$H$_2$, HCN, and OH lies in channel 3 where blending between sources A and B is significant, we fitted these species using spectra from the combined A+B aperture. For all species in channels 1 and 2, we fitted the spectra of source A and B independently. We show an example of the combined best-fit model to source A+B for each molecule in Figure \ref{fig:total_model_spectra_AB}; the fits to the remaining molecules and sources are shown in the appendix in Figures \ref{fig:total_model_spectra_A_H2O} to \ref{fig:total_model_spectra_B_H2O}. 

For each model, we obtain a best-fit $N$, $T_\mathrm{ex}$, and $R$, and also calculate the total number of molecules $\mathcal{N}=\pi*R_\mathrm{best}^2*N_\mathrm{best}$ and a column density ratio with respect to H$_2$. The column densities of H$_2$ for each source have been obtained with a rotational diagram analysis by \cite{Gieser2023}. Two temperature components are used in the rotational diagram fitting, and we use the column density of the cooler $T_\mathrm{rot}\sim300$ K H$_2$ for our abundance calculations (see Appendix of \citealt{Gieser2023}). We note that this is warmer than the best-fit temperatures for CO$_2$, C$_2$H$_2$, and HCN, and cooler than the temperatures for the H$_2$O  absorption, and thus the warm H$_2$ may not trace the same reservoir of gas. The H$_2$ emission from IRAS 23385+6053 clearly traces the outflows as well as the protostar positions (See Figure \ref{fig:apertures} left panel), and thus our inferred abundances may be higher in reality if only a fraction of the H$_2$ emission originates from the quiescent disk and envelope. However, \cite{Gieser2023} also determine H$_2$ column densities using the mm continuum emission and an assumed gas-to-dust ratio, and find column densities consistent with the warm H$_2$ within a factor of a few. The best-fit parameters of our models and estimated column density ratios are summarized in Table \ref{tab:best_fit_values}.

Confidence interval contours in the $N$-$T_\mathrm{ex}$ plane for each fit are overlaid on maps of normalized $\chi^2$ in Figures \ref{fig:chi2maps_sourceAB} to \ref{fig:chi2maps_sourceN} of the appendix. The confidence contours can be complex but typically show one of the following behaviours. 1) The emission is optically thick and the confidence contours are banana-shaped, indicating that $R$ is well-constrained, while $N$ and $T_\mathrm{ex}$ have a non-linear degeneracy. 2) The emission is optically thin and the confidence contours are tongue-shaped, indicating $T_\mathrm{ex}$ is well-constrained, but $R$ and $N$ are fully degenerate, and only lower and upper limits respectively can be determined. However, in this case the total number of molecules $\mathcal{N}$ is still well-constrained. 3) The $S/N$ of the molecular features is low, resulting in confidence intervals showing a combination of behaviours 1 and 2. Here, solutions with either cooler and more optically thick emission or hotter and more optically thin emission are degenerate.  

\begin{figure*}
    \centering
    \includegraphics[width=\textwidth]{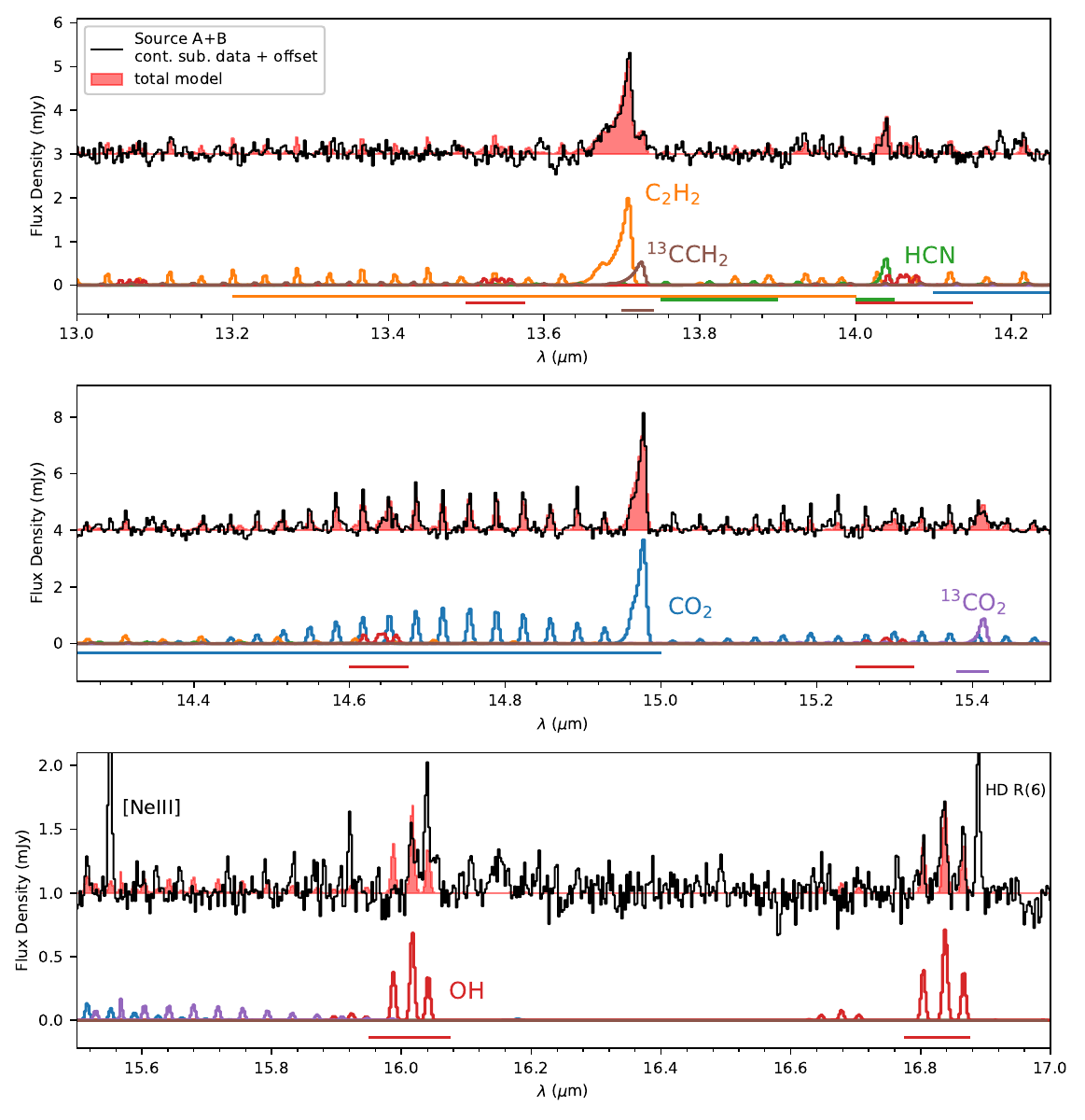}
    \caption{Combined best-fit LTE model for $^{12}$CO$_{2}$, $^{13}$CO$_{2}$, C$_{2}$H$_{2}$, $^{13}$CCH$_{2}$, HCN, and OH emission in source A+B (red shaded region) overlaid on the continuum subtracted data with an arbitrary flux offset (black). Prominent emission lines are labelled. Best-fit results for individual molecules are shown as the coloured lines below, with the horizontal bars indicating the regions of spectrum used for fitting. Extinction by CO$_2$ ice has been included in the models, which suppresses the flux from the CO$_2$ gas $P$-branch (See section \ref{ssec:slab_models}).}
    \label{fig:total_model_spectra_AB}
\end{figure*}

\begin{table*}
\small
\centering
\caption{LTE slab model best-fit parameters}
\label{tab:best_fit_values}
\begin{tabular}{cccccc}
\hline \hline
Species         & $N$ (cm$^{-2}$)         & [$N/N$ H$_2$]$~^a$           & $T$ (K)  & $R$ (au) & $\mathcal{N}$ (mol.)\\
\multicolumn{6}{c}{source A+B} \\
\hline
$^{12}$CO$_{2}$    & $1.7${\raisebox{0.5ex}{\tiny$^{+1.3}_{-0.3}$}} ($17$) & $2.0${\raisebox{0.5ex}{\tiny$^{+1.6}_{-0.3}$}} ($-7$) & $1.2${\raisebox{0.5ex}{\tiny$^{+0.1}_{-0.1}$}} ($2$) & $1.8${\raisebox{0.5ex}{\tiny$^{+1.1}_{-0.3}$}} ($2$) & $3.7$ $(48)$ \\ 
 $^{13}$CO$_{2}~^b$    &  $<1.2$ ($16$) & $< 1.4$ ($-8$) & - & - & $2.6$ $(47)$ \\ 
 C$_{2}$H$_{2}$     &  $7.6${\raisebox{0.5ex}{\tiny$^{+3.6}_{-3.6}$}} ($16$) &  $9.0${\raisebox{0.5ex}{\tiny$^{+4.6}_{-4.6}$}} ($-8$) &  $2.4${\raisebox{0.5ex}{\tiny$^{+0.3}_{-0.3}$}} ($2$) &  $1.4${\raisebox{0.5ex}{\tiny$^{+0.4}_{-0.2}$}} ($1$) &  $1.0$ $(46)$ \\  
 $^{13}$CCH$_{2}~^c$   &  $<8.5$ ($15$) &  $<1.0$ ($-8$) & - & - &  $1.1$ $(45)$ \\  
HCN                & $< 2.7\,(19)$ & $< 3.1\,(-5)$ & $< 1.0\,(2)$ & $> 4.0\,(2)$ & $2.4$ $(49)$ \\ 
OH$~^d$             & $< 7.9\,(17)$ & $< 9.4\,(-7)$ & $1.8${\raisebox{0.5ex}{\tiny$^{+0.6}_{-0.5}$}} ($3$) & $> 2.0\,(0)$ & $4.0$ $(44)$ \\  
\multicolumn{6}{c}{source A} \\
\hline
CO$~^e$       & $< 5.6\,(18)$ & $< 6.3\,(-6)$ & $2.4${\raisebox{0.5ex}{\tiny$^{+0.7}_{-1.0}$}} ($3$) & $> 8.8\,(-2)$ & $4.7$ $(42)$ \\
\multicolumn{6}{c}{source B} \\
\hline
H$_{2}$O$~^f$& $8.2${\raisebox{0.5ex}{\tiny$^{+77.3}_{-5.7}$}} ($18$) & $1.0${\raisebox{0.5ex}{\tiny$^{+9.5}_{-0.7}$}} ($-5$) & $5.5${\raisebox{0.5ex}{\tiny$^{+5.0}_{-3.0}$}} ($2$) & - & - \\
\multicolumn{6}{c}{source S1} \\
\hline
CO$~^e$       &$< 5.4\,(17)$ & $< 5.4\,(-7)$ & $1.9${\raisebox{0.5ex}{\tiny$^{+0.3}_{-0.2}$}} ($3$) & $> 2.2\,(-1)$ & $1.1$ $(43)$ \\ 
OH                 & $< 3.8\,(18)$ & $< 3.8\,(-6)$ & $1.3${\raisebox{0.5ex}{\tiny$^{+1.6}_{-0.3}$}} ($3$) & $> 1.2\,(0)$ & $1.4$ $(45)$ \\ 
\hline
\end{tabular}
\\ All numbers are written in the format $a \,(b) = a \times 10^b$, with 1 sigma confidence intervals where appropriate.
\\ $^a$ See discussion of uncertainties and caveats on the column density ratios in Sections \ref{ssec:lte_slab_fits} and \ref{ssec:chemistry}.
\\$^b$ Fitting performed with $T$ and $R$ fixed to best-fit values for $^{12}$CO$_2$. 
\\$^c$ Fitting performed with $T$ and $R$ fixed to best-fit values for C$_2$H$_2$.
\\$^d$ Highly non-LTE behaviour likely, see Section \ref{ssec:OH_disc}.
\\$^e$ Best-fit temperature can be overestimated, see Section \ref{ssec:CO_disc}.
\\$^f$ Seen in absorption.
\end{table*}

\subsubsection{Source A and B}

The high $S/N$ of the CO$_2$ and C$_2$H$_2$ emission allows the parameters of the LTE model fits to be particularly well constrained. For the CO$_2$ emission we fitted only the $Q$ and $R$-branch.  The width of the Q-branch and distribution of the R-branch transitions is particularly sensitive to the excitation temperature in the model fits. We do not fit the $P$-branch as its $S/N$ is lower and its flux is thus more sensitive to the continuum determination and correction from ice absorption. For C$_2$H$_2$, only the $Q$-branch is clearly detected and thus used for fitting. The slab model fits for both species indicate optically thick emission, with well-constrained temperatures of $120 \pm 10$ K and $180 \pm 30$ K respectively. The best-fit emitting area radii of  CO$_2$ ($180^{+110}_{-30}$ au) is larger than that of C$_2$H$_2$ ($14^{+14}_{-2}$ au) by a factor of $\sim$13. The emission for HCN has a similar best-fit temperature of $\sim 60$ K, but is optically thin and has a poorly constrained emitting area radius of $> 400$ au. 

Optically thin emission from the isotopologues $^{13}$CO$_2$ and $^{13}$CCH$_2$ is  tentatively detected, but due to the low $S/N$ of these features, the exact column densities and temperatures can not be well determined when fit without any constraints.
We thus perform the fitting for $^{13}$CO$_2$ and $^{13}$CCH$_2$ with the model $T_\mathrm{ex}$ and $R$ fixed to the best-fit values from CO$_2$ and C$_2$H$_2$ respectively,  and report the resulting column densities as upper limits. The $^{13}$C/$^{12}$C column density ratio is thus $>14$ using the CO$_2$ and $^{13}$CO$_2$ fits and $>9$ using the C$_2$H$_2$ and $^{13}$CCH$_2$ fits. This ratio is consistent with what is expected from the ISM abundance at a galactocentric distance of 11 kpc of $87 \pm 15$ from the models of \cite{Milam2005}, but higher S/N data is needed to place better constraints on the $^{13}$C/$^{12}$C ratio. 

Faint OH emission is detected from both sources and is blended with other molecules except for the groups of transitions at 16.0 and 16.8 micron. The OH emission appears optically thin and has a high $T_\mathrm{ex}$ of $1800^{+600}_{-500}$ K, as expected from its highly non-LTE formation process (see Section \ref{ssec:OH_disc}).

For molecules at shorter wavelengths where source A and B can be spatially resolved, there are stark differences between the two spectra. Emission from the $P$-branch of CO is only detected towards source A. The best-fit solutions suggests optically thin emission with $N < 5.6 \times 10^{18}$ cm$^{-2}$ and $T_\mathrm{ex}$ of $2400^{+700}_{-1000}$ K, but solutions with optically thick emission and relatively cooler temperatures of 500-1000 K are also plausible at a 2$\sigma$ level (see Section \ref{ssec:CO_disc}). H$_2$O is detected in emission from source A but in absorption from source B. The rovibrational emission towards source A is detected at low $S/N$, and the fitting is further complicated by the difficulty of determining the continuum for such faint features. In Figure \ref{fig:total_model_spectra_A_H2O} we thus only show a representative model with a by-eye fit to illustrate the detection. Here, the emitting area radius is set to the best-fit for CO$_2$ of 180 au, a column density of $10^{14}$ cm$^{-2}$, and excitation temperature of 500 K. A handful of pure rotational transitions of H$_2$O in source A are also present at 16-17 $\mu$, though at this wavelength blending with source B is strong. Here, these transitions are consistent with optically thin emission at a temperature of $\sim 350$ K but and column densities of $<10^{18}$ cm$^{-2}$. In source B, the absorption model has a best-fit temperature of $550^{+500}_{-300}$ K and column density of $8.2^{+77.3}_{-5.7} \times 10^{18}$ cm$^{-2}$, which is not well constrained due to the low $S/N$ and uncertainty in the continuum determination for H$_2$O transitions deep within ice absorption features.

\subsubsection{H$_2$ hotspots}

In S1, we only detect CO $P$-branch, OH, and H$_2$ emission. Towards the other H$_2$ hotspots, no other molecular emission is detected. Notably, no CO$_2$ emission is detected, as was seen with ISO Short Wavelength Spectrometer (SWS) towards the Orion-IRC2/KL shock positions \citep{Boonman2003c}. The CO emission appears to be optically thin and consistent with a well-constrained temperature of $1900^{+300}_{-200}$ K. The best-fit emitting radius is larger than a few au, but degenerate with column density. The OH emission has a best-fit solution with $\sim 1300$ K optically thick emission, but higher temperatures and an optically thin solution are also possible, as is the case for source A+B.

\section{Discussion}
\label{sec:disc}

Here, we briefly discuss some of the options for the origin of the molecular emission, before discussing each molecule in turn, as they are likely tracing distinct components in IRAS 23385+6053.

\subsection{Origin of the emission}
\label{ssec:origins_disc}

Identification of the origin of the emission is complicated by several factors. The spatial resolution of MIRI/MRS at the distance to IRAS 23385+6053 ranges from 1300-5000 au, and thus the extraction apertures for our spectra (Figure \ref{fig:apertures}) may contain emission from a diverse range of environments, including the hot inner disk, warm and colder envelope, irradiated outflow cavity, and gas in outflows with both $C$ and $J$ type shocks. Furthermore, the emission from sources A and B begin to overlap in channel 3, and are entirely blended in channel 4. The fits to source A+B thus indicate the properties of molecular emission in a large region, probably containing at least two disks.

The assumption of LTE at the gas temperature $T_\mathrm{kin}$ is likely not appropriate for several of the detected species, as this requires that collisions are the dominant excitation mechanism, and that the density of molecular hydrogen exceeds the critical density for the observed transitions, which are of order 10$^{12}$ cm$^{-3}$ for rovibrational transitions \citep{Bruderer2015}. While the typical densities in disks may be large enough for collisional excitation, rovibrational molecular emission has also been observed in the much more tenuous gas of protostellar outflows \citep[e.g.][]{Sonnentrucker2006,Sonnentrucker2007,Tappe2012}. Here, radiative pumping from the mid-IR thermal dust continuum or UV emission produced in shocks may instead be responsible for the excitation, and a non-LTE model including these effects would be more appropriate. If mid-IR pumping is the dominant excitation mechanism, the best-fit LTE model temperature may reflect the temperature of the radiation field, which likely originates from the thermal dust continuum. Furthermore, pumping can cause the column density to be overestimated by several orders of magnitude. This effect has recently been identified in observations of SO$_2$ emission from protostars at both mid-IR and sub-mm wavelengths, which confirm the presence of strong mid-IR pumping \citep{vanGelder2023}. However, in general the strength and shape of the local radiation fields are highly uncertain, and depend on the local extinction and source geometry. Even if molecular emission does originate in the disk, large gradients in temperature and density should be present, and the assumption of a single temperature may not be valid. Nonetheless, the LTE slab models provide a good fit to the observations, except for the case of H$_2$O. We discuss in the following sections the case of each molecule in detail, and note where non-LTE effects may be important.

\subsection{Simple organics - CO$_2$, C$_2$H$_2$, HCN}
\label{ssec:simple_organics_disc}

Emission from the simple organic molecules CO$_2$, C$_2$H$_2$, and HCN has often been detected in absorption towards high mass protostars in the hot core phase \citep[e.g][]{Lahuis2000,Boonman2003c,Barr2020}. However, IRAS 23385+6053 is likely in an earlier stage of evolution, as indicated by the lack of a compact HII region or free-free emission \citep{Molinari1998}, and the low number of detected COMs associated with high temperature chemistry and ice sublimation \citep{Cesaroni2019,Gieser2021}. The general lack of absorption, which requires colder absorbing gas in front of a warm continuum along the line of sight, is thus not surprising. The similar temperatures (100-200 K) and emitting areas (10s to 100s of au) of the LTE model fits suggest a common origin for the simple organics. For source A+B, the CO$_2$ and C$_2$H$_2$ emission is optically thick while the HCN emission is optically thin, which is consistent with models that predict the strongest HCN transitions become optically thick at about an order of magnitude higher column density than C$_2$H$_2$ \citep{Lahuis2000}. 

The simple organics may be tracing a variety of components of a young protostellar system. Previous {\it Spitzer} observations of protostars have found emission from these species to be associated with the outflows. In {\it Spitzer} Infrared Spectrograph observations of the Cepheus A star-forming region ($d=690$ pc), \cite{Sonnentrucker2006,Sonnentrucker2007} found spatially extended ($\sim21000$ au) CO$_2$ and C$_2$H$_2$ emission with temperatures of 50-200 K and column densities of $10^{17}$ to $10^{19}$ cm$^{-2}$.
In the Orion-IRc2/KL ($d\sim 450$ pc) region, \cite{Boonman2003c} similarly detected CO$_2$, C$_2$H$_2$, and HCN emission from H$_2$ bright shocked regions. In both cases, the low densities of the outflow regions relative to the critical densities of the simple organics supported mid-IR pumping by the local dust as the dominant excitation mechanism for the observed emission. However, while \cite{Sonnentrucker2006,Sonnentrucker2007} find a strong correlation between the spatial extent of the simple organics and extended H$_2$ emission, in IRAS 23385+6053 these molecules are only detected towards source A and B, have emitting area radii $<200 $ au, and are not seen at the outflow knot positions (c.f. Figure \ref{fig:spectra_overview} and Table \ref{tab:detected_molecules}). This would suggest that the simple organics do not primarily originate in the shocks, and are instead more likely to be associated with a protostellar disk and/or dense inner envelope. The prior detection towards source A of large scale gas motions resembling combined keplerian rotation and infall onto a nearly face on disk supports the disk scenario \citep{Cesaroni2019}. 

The large emitting area radii ($>10$s of au), temperatures of 100-200 K, and lack of strong water emission argues against the emission arising from the hot inner disk, as has typically been found in the more evolved T Tauris objects \citep{Carr2011}. Instead, the emission may come from the extended warm disk surface and dense inner envelope that would be expected around young high mass protostars. The emitting radius for CO$_2$ of $\sim180$ au is much larger than a typical T Tauri star disk, but is consistent with interferometric observations of high mass protostars that suggest disks of hundreds to thousands of au in radius \citep[e.g.][]{Johnston2015,Maud2019}. Models of high-mass embedded disks by \cite{Nazari2022} (see also \citealt{vanDishoeck2023}) indeed show typical temperatures of 150-200 K at radii of $\sim100-200$ au. The dense environment of the disk should exceed the critical densities of the simple organic molecules, and in this case the assumption of LTE may be reasonable. However, we cannot rule out a contribution from mid-IR pumping from the dust in the disk or outflows. The lack of mid-IR continuum emission towards the H$_2$ knot positions suggests that pumping would only be important close to source A and B, but more modelling of the location radiation field is required to evaluate this scenario fully and is beyond the scope of this paper. 
 
\subsection{H$_2$O}
\label{ssec:H2O_disc}

Similar to the case of the simple organics, H$_2$O is likely not associated with large scale molecular outflows, as the emission and absorption is only detected towards sources A and B, and none of the H$_2$ knot positions (c.f Figure \ref{fig:spectra_overview} and Table \ref{tab:detected_molecules}). This is in stark contrast to the ubiquitous detection of H$_2$O far-IR and sub-mm lines in shocks and outflows \citep{vanDishoeck2021}. Mid-IR H$_2$O emission in protostars has not been well characterized previously owing to the relative lack of sensitivity of previous space missions compared to JWST. In the low mass protostars of the {\it Spitzer} c2d survey, H$_2$O emission was rarely detected (12/43 sources), and for all but one source, only in the more evolved class I stage \citep{Lahuis2010}. For these more evolved sources, the H$_2$O emission appeared to have similar properties to Class II disks, with hot temperatures and small emitting radii. In the sole mid-IR detection towards a Class 0 source, the H$_2$O emission appeared cold ($T_\mathrm{ex}\sim170$ K) and was suggested to originate from an accretion shock onto the disk \citep{Watson2007}, although this was later disputed by an analysis of far-IR H$_2$O emission with {\it Herschel} \cite{Herczeg2012}. In source A, the parameters of the H$_2$O emission are not well determined, but the temperature of our representative model of $\sim$ 500 K would be consistent with an origin in the disk surface.

In previous observations of high mass protostars, H$_2$O has typically been detected in absorption \citep{Boonman2003a,Boonman2003b,Indriolo2020,Barr2022}. Two non-mutually-exclusive possible scenarios involve a disk origin of the absorption and have been proposed for previous observations of protostars. In the first, absorption occurs in the disk atmosphere against the continuum from the midplane, which is viscously heated by accretion \citep[e.g][]{Barr2022}. In the second, the disk is viewed nearly edge on relative to the observer, and the absorption occurs against the continuum from the hot inner disk on a line of sight intersecting the disk \citep[e.g., ][]{Lahuis2006,Knez2009}. A difference in inclination is an attractive possibility for explaining why we see H$_2$O in emission in source A but in absorption in source B. However, a nearly edge-on disk would also be expected to produce strong absorption features from CO$_2$, C$_2$H$_2$, and HCN which we do not detect \citep{Lahuis2006,Knez2009}. A third possibility is that the H$_2$O absorption originates in a dense inner disk wind against the continuum from the disk, but this may require a very specific inclination angle of the disk which matches the opening angle of the wind. We propose that the most likely scenario is that the disks in source A and B are both viewed at a low to moderate inclination, but that the viscous heating of the mid-plane is stronger in source B due to a much higher accretion rate, resulting in absorption.  In source A, the weak H$_2$O emission then traces the warm inner disk, similar to what has been observed for Class II disks \citep{Carr2011}.

\subsection{CO}
\label{ssec:CO_disc}

\begin{figure}[htb]
    \centering
    \includegraphics[width=0.5\textwidth]{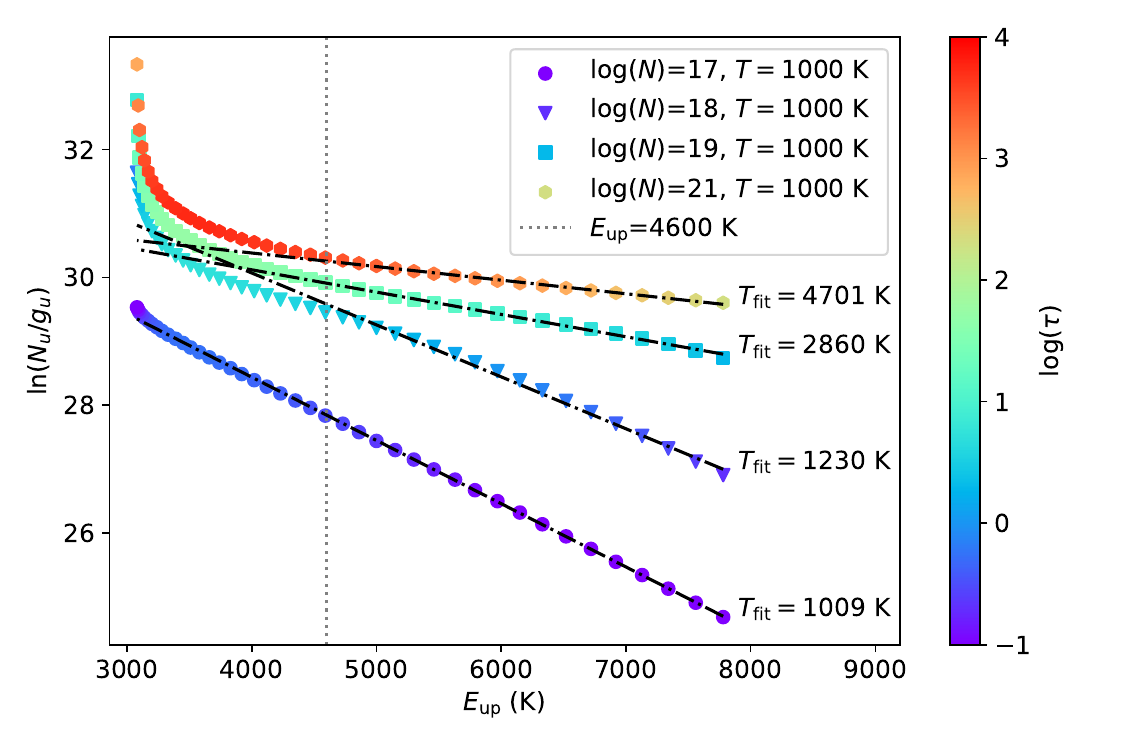}
    \caption{Rotational diagrams of CO v=1-0 $P$-branch transitions produced using our LTE slab model for four different column densities at a temperature of $T=1000$ K; the colour of each symbol indicates the line optical depth. The dotted line at $E_\mathrm{up}=4600$ K is the energy above which transitions fall shortwards of the MIRI/MRS blue wavelength limit of 4.9 $\mu$m. For each model, the rotational temperature derived from a linear fit to transitions in the MIRI range (i.e. above $E_\mathrm{up}=4600$ K) is shown by the black lines.}
    \label{fig:co_trot_miri}
\end{figure}

In MIRI/MRS observations, the detection of any CO emission indicates highly energetic excitation conditions, as the shortest wavelength cutoff of 4.9 $\mu$m limits us to the detection of $P$-branch transitions with high upper energy levels ($E_\mathrm{up} > 4600$ K, $J_\mathrm{up}> 24$). The best-fit LTE CO models for source A and source S1 appear to indicate optically thin emission with high temperatures $T > 1900$ K. However, even with the simplifying assumption of LTE, it may not be possible to robustly determine the gas temperature due to the combination of optical depth effects and the lack of lower energy transitions \citep{Herczeg2011}. To illustrate this, we show in Figure \ref{fig:co_trot_miri} synthetic rotational diagrams of the CO $v=1-0$ $P$-branch produced using a 1000 K LTE slab model at a variety of column densities, with the optical depth of each line indicated by the colour scale. At $E_\mathrm{up}\gtrsim4600$ K (dotted line), the $v=1-0$ $P$-branch transitions lie within the MIRI/MRS range at $\lambda>4.9~\mu$m. A linear fit to the MIRI/MRS range transitions recovers the true temperature of 1000 K for a very optically thin model ($N=10^{17}$ cm$^{-2}$), but is increasingly biased towards higher temperatures as the emission becomes optically thick. While our LTE models can account for optical depth effects, the curvature in the rotational diagram is only easily recognized in the lower upper-energy transitions located at $<4.9~ \mu$m. This effect similarly holds for the $v=2-1$ $P$-branch transitions in the MIRI range, which have even higher upper energy levels of $> 7000$ K. Observations with instruments covering shorter wavelengths (e.g. JWST/NIRSpec or VLT/CRIRES+) are therefore required for accurate modelling.

With this in mind, we can only conclude reliably that the observed CO emission in sources A and S1 originates from gas with a high excitation temperature, $T\gtrsim1000$ K. What could explain this CO emission (and the non-detection of CO emission in source B)? CO rovibrational emission from protostars has been studied in detail from ground based high spectral resolution observations. Analysis of the emission line velocity profiles typically indicates that the emission originates from a broad and hot component ($T\sim 1000$ K) in the inner edge of the disk -  similar to what is commonly observed for Class II disks \citep{Salyk2011}  - or within a narrow and cooler ($T\sim 300$ K), component associated with a disk wind or outflow shocks \citep{Pontoppidan2002,Herczeg2011}. Assuming that LTE holds, an inner disk-origin of the CO emission seems most reasonable for source A, and would be consistent with the other molecule detections. However, CO can be significantly pumped by strong mid-IR radiation \citep{Blake2004,Thi2001}, and an outflow origin cannot be ruled out. An outflow origin is more likely the case for the H$_2$ knot in source S1. Here, strong outflow activity is suggested by the high H$_2$ line velocities, but we do not detect mid-IR continuum emission, nor any of the H$_2$O or simple organics seen towards source A B that suggest a disk. For source B, the reason for the non-detection of CO is unclear, as a disk is likely present given the significant H$_2$O absorption, and there is significant outflow activity within this aperture as well. High spectral resolution studies of CO have shown that the line profiles for some sources can show a complex velocity structure with a combination of emission and absorption from the disk and disk winds \citep{Herczeg2011,Thi2010}. At low spectral resolution, blending of these features may smooth out the line profiles to a very low flux, making the CO difficult to detect. This scenario would be more likely if the H$_2$O absorption indeed originates from a disk wind. Alternatively, the emitting region for CO may be much smaller in source B, and thus the emission would simply be below our detection threshold.

\subsection{OH}
\label{ssec:OH_disc}

The OH transitions detected towards sources A+B and S1 are best fit by optically thin LTE models with high excitation temperatures of $T_\mathrm{ex} > 1300$ K. However, the detected transitions from 13-17 $\mu$m have very high upper levels  ($E_\mathrm{up} \sim 8000 - 10000$ K, $N=21-17$) as well as large Einstein A coefficients ($A_{ij} \sim 100$ s$^{-1}$), indicating that collisional excitation is not plausible as an excitation mechanism and that strong non-LTE processes are at play. Mid-IR OH emission has previously been detected in environments with strong UV irradiation, such as shocks from protostellar outflows \citep{Tappe2008,Tappe2012} and the surface layers of Class II disks \citep{Carr2014}. Here, OH transitions in extremely high rotational states ($E_\mathrm{up} \sim 40000$ K) but low lying vibrational states are observed. These transitions can be explained by the photodissociation of H$_2$O by UV photons in the $\tilde{B}=114-143$ nm absorption band, which produces OH in extremely hot rotational states, followed by `prompt' emission of photons associated with a cascade of pure rotational transitions to the ground state \citep{Tabone2021,Zannese2023}. However, these prompt transitions lie primarily at $9-11~\mu$m and are not detected in our observations. Furthermore, the rotational lines of OH are split into a quadruplet by spin-orbit coupling and $\Lambda$-doubling, and for prompt OH emission, a strong asymmetry is expected between the components, but we do not observe this \citep{TaboneinPrep}. A different non-LTE mechanism is the formation pumping of OH to highly excited rotational and vibrational states as the result of the O + H$_2$ $\rightarrow$ OH + H reaction \citep{Liu2000}, which can produce the emission longwards of $\sim13$ $\mu$m we observe. Further non-LTE modelling is required to understand the excitation mechanism, but is beyond the scope of this paper. The lack of detection of OH towards the H$_2$ knots S2 and S3 could be explained by weaker shock activity at these locations, resulting in OH emission below our detection limits.

\subsection{Chemistry}
\label{ssec:chemistry}

Here, we comment on the chemistry for the simple organics CO$_2$, C$_2$H$_2$, and HCN, which have relatively precise LTE model fits. We caution that if significant mid-IR pumping is present, this would bias our LTE model fits towards higher column densities and abundances than are actually observed, and the excitation temperature would reflect the temperature of the radiation field rather than the kinetic temperature of the gas. The H$_2$ column densities used for our abundance determination also likely traces gas in the outflows as well as the disk and envelope, which would bias out abundances towards lower values.

We proceed here assuming that emission from these molecules originates in the disk and is indeed in LTE. What then is the origin of the molecules themselves? The presence of deep CO$_2$ ice absorption features at $\sim 14.9$ $\mu$m clearly indicates the presence of abundant CO$_2$ on the grain mantles along the line of sight in the larger scale envelopes. The dust temperature towards sources A and B estimated from a black-body fit of $\sim 150$ K exceeds the sublimation temperature of CO$_2$ of $\sim 55$ K, indicating thermal desorption closer to the protostars. The detection of [FeII], [SI] towards source A and B \citep{Gieser2023} implies strong shocks capable of sputtering ices from the grain mantle are also present. Either or both mechanisms could release CO$_2$ into the gas phase, and a similar scenario could apply for the C$_2$H$_2$ and HCN emission, although both of these molecules have yet to be detected in ices. 

Gas-phase chemistry may also contribute to the formation of the observed molecules. This may particularly be likely for CO$_2$, which has enhanced production by the reaction OH + CO $\rightarrow$ CO$_2$ + H at temperatures of 100-250 K. Above this temperature range, Oxygen is instead driven into H$_2$O by OH + H$_2$ $\rightarrow$  H$_2$O + H, and CO$_2$ enhancement is not expected \citep{vanDishoeck2023}. For HCN and C$_2$H$_2$, gas phase production can occur efficiently at temperatures $>200$ K via the CN + H$_2$ $\rightarrow$ HCN + H reaction for HCN, and through multiple pathways for C$_2$H$_2$ \citep{Doty2002}. However, our LTE model fits find excitation temperatures $<200$ K for both molecules, suggesting that gas phase formation is not important.

In the context of hot core observations of gas-phase absorption, the abundance of C$_2$H$_2$ and HCN has been found to be correlated with $T_\mathrm{ex}$ \citep{Lahuis2000}, while CO$_2$ shows no such trends \citep{Boonman2003d}. The abundance of HCN is only constrained by our modelling to $<10^{-5}$, but C$_2$H$_2$ and CO$_2$ abundances of $\sim10^{-7}$ are consistent with hot cores with $T_\mathrm{ex}\sim200$ K. The hot core chemistry models of \cite{Lahuis2007} provide predictions for the CO$_2$, C$_2$H$_2$, and HCN abundance over a range of excitation temperatures and X-ray fluxes. Comparing our CO$_2$ model fit results with Figure 3 of \cite{Lahuis2007}, at a $T_\mathrm{ex}$ of $\sim 120$ K the CO$_2$ abundance is very sensitive to X-ray flux. However, our best-fit abundance of $\sim2\times10^{-7}$ is possible for some models with lower X-ray flux ($<1$ erg s$^{-1}$ cm$^{-2}$). For C$_2$H$_2$ at $\sim 180$ K, very low abundances of $<10^{-9}$ are expected, but we have $\sim3\times10^{-7}$. Deeper JWST searches for both HCN and C$_2$H$_2$ ices are warranted, which could mitigate the C$_2$H$_2$ discrepancy.

\label{ssec:chem_disc}

\section{Summary and future outlook}
\label{sec:summary}

In Figure \ref{fig:cartoon}, we summarize the proposed origins of the different molecular emission components observed towards IRAS 23385+6053. Taken together, the detected molecular species are largely consistent with an origin in a young disk and inner envelope, with a possible contribution from a dense inner disk wind.
The major caveats to this picture are the limited spatial resolution at the large distance to the source, and the possibility of non-LTE effects, which we discuss in Section \ref{ssec:origins_disc}. With this in mind, the interpretation of our models is as follows:

\begin{itemize}
    \item In the blended spectrum from source A+B, warm $120-180$ K emission of CO$_2$, C$_2$H$_2$, and HCN is seen. The origin is most likely predominantly in the warm disk surface or dense inner envelope, but some contribution from the base of an outflow within the JWST beam cannot be ruled out. These species are not detected towards the shock positions in source S1-S3.
    \item H$_2$O is only detected towards source A and B, and not towards any shock positions, suggesting a disk rather than a large scale outflow origin. In source B, strong accretion heating in the disk midplane can produce a temperature inversion, resulting in hot $250-1050$ K H$_2$O absorption. Alternatively, the absorption may occur in a dense inner disk wind, though this requires a very specific geometry. In source A, the accretion heating is weaker, and some faint H$_2$O emission is detected, which may originate in the inner disk.

\end{itemize}

\vfill
\noindent
\begin{minipage}{1.0\textwidth}
  \strut\newline
  \centering
  \includegraphics[width=0.8\textwidth]{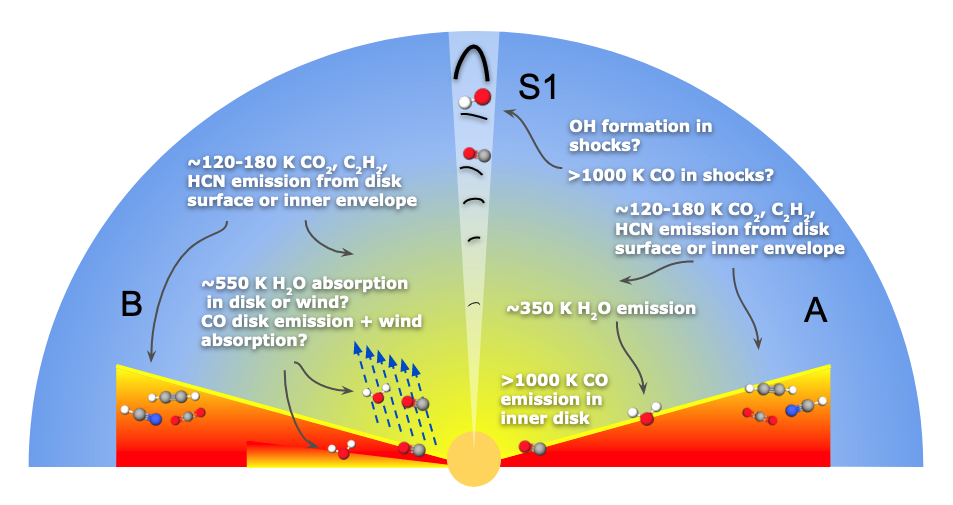}
  \captionof{figure}{Schematic of our proposed origins for the molecular emission and absorption observed towards source A (right half), B (left half), and S1 (outflow shocks, centre) in IRAS 23385+6053.}\label{fig:cartoon}
\end{minipage}

\begin{itemize}
    \item CO tracing an inner disk and/or outflow shocks is detected towards source A and S1. Only high-$J$ $P$-branch states are detectable in the MIRI/MRS range of 4.9-28.3 $\mu$m, which suggests high temperatures $T>1000$ K. Observations at shorter wavelengths covering the lower J lines are needed to determine the temperature if the emission is optically thick. CO is not detected towards source B, possibly due to absorption in a disk wind being filled in by emission from the disk. 
    \item Highly non-LTE OH emission is detected towards source A, B, and S1, but not S2 or S3. Photo-dissociation or chemical formation pumping is likely responsible for the emission in high rotational states observed, although we do not detect the prompt OH transitions at $9-11$ $\mu$m or $\Lambda$-doublet asymmetry associated with photo-dissociation.
\end{itemize}
    
There are many opportunities for complementary observations and modelling to improve the exploitation of MIRI/MRS data. For complex and distant high mass star-forming regions, spatial resolution is still a limiting factor. Long-baseline observations with ALMA can approach the scale of the disk, though these are not possible for IRAS 23385+6053 given its high latitude relative to the ALMA observatory. At shorter wavelengths, ground based observations of CO and H$_2$O with high spectral resolution can provide valuable kinematic information for distinguishing between a disk and outflow origin, although sources like IRAS 23385+6053 are too faint for 8m class telescopes. Observations with JWST/NIRSpec would capture the full CO rovibrational band and a variety of accretion tracing HI lines accessible in the near-IR. On the modelling front, non-LTE analysis will be important for molecular emission detected in the high energy and low density environments of protostellar outflows. In particular, analysis of prompt OH emission produced by photo-dissociation of H$_2$O can provide valuable insights into the local UV fields and density of H$_2$O. 

The unprecedented sensitivity of JWST is opening a new window into high mass star formation 
allowing extremely faint and distant sources to be detected and characterized for the first time. With the imminent arrival of additional MIRI/MRS observations of high mass protostars, our understanding of how a very young object such as IRAS 23385+6053 fits into the broader evolution will greatly improve. 

\newpage
\begin{acknowledgements}

We would like to thank the anonymous referee for their very constructive discussions on this paper. 

This work is based on observations made with the NASA/ESA/CSA James Webb Space Telescope. The data were obtained from the Mikulski Archive for Space Telescopes at the Space Telescope Science Institute, which is operated by the Association of Universities for Research in Astronomy, Inc., under NASA contract NAS 5-03127 for JWST. These observations are associated with programme 1290. 

Astrochemistry in Leiden is supported by funding from the European
Research Council (ERC) under the European Union’s Horizon 2020
research and innovation programmeme (grant agreement No. 101019751
MOLDISK), by the Netherlands Research School for Astronomy (NOVA),
and by grant TOP-1 614.001.751 from the Dutch
Research Council (NWO).
\end{acknowledgements}

\bibliographystyle{aa} 
\bibliography{references.bib} 

\begin{appendix}

\section{Thermal continuum and baseline modelling}
\label{sec:app_baseline}

\begin{figure*}[htb]
    \centering
    \includegraphics[width=0.8\textwidth]{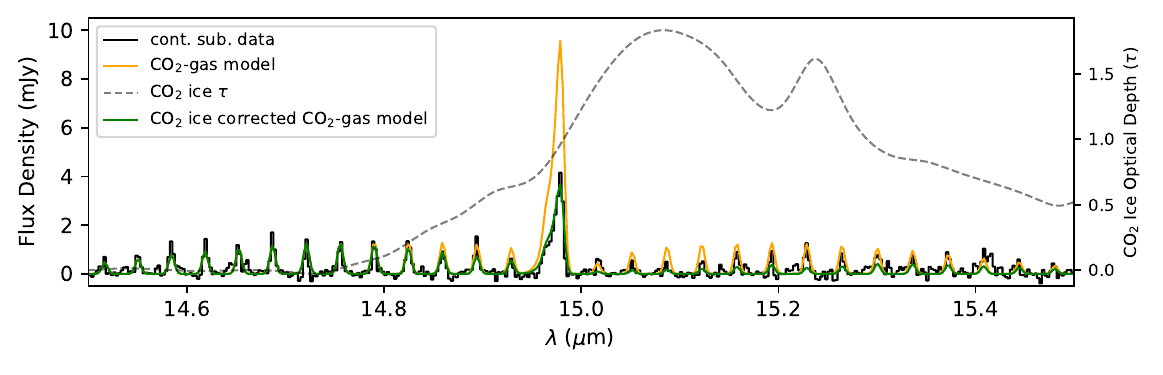}
    \caption{Correction for differential extinction by CO$_2$ ice of the CO$_2$ gas emission features. The blue line shows the continuum subtracted data with the CO$_2$ ice feature removed. The best-fit LTE model of CO$_2$ with extinction correction, but without any correction for ice absorption (orange) overproduces the $Q$ and $P$-branch emission. The estimated optical depth of the CO$_2$ ice feature (dashed grey line) is applied as a scaling factor to the same best-fit model (green curve), which significantly improves the fit to the Q-branch,  though the P-branch features are now somewhat under-produced. 
    }
    \label{fig:CO2_ice_corr}
\end{figure*}

Here, we show examples of the local continuum modelling procedure applied to our data and additional extinction corrections applied to our gas-phase emission models. In Figure \ref{fig:cont_spline}, examples of cubic spline fits to the local continuum at shown by the black points and solid lines overlaid on the data for source A in green. Additionally, the thermal dust continuum estimated with a cubic spline fit (dashed line) to the red data points is indicated. These estimates of the thermal dust continuum are used to calculate the wavelength dependent extinction of the ice features $\tau_\mathrm{ice}= -\log(F_\mathrm{dust}/F_\mathrm{continuum})$. The wavelength dependent extinction is applied as a correction factor $\exp (-\tau_\mathrm{ice})$ to our LTE models for emission from molecular species overlapping with the ice features, as demonstrated in Figure \ref{fig:CO2_ice_corr} for the gas of gas-phase CO$_2$. The effect in this case is to suppress the emission from the $Q-$ and $P-$ branch.

\begin{minipage}{1.0\textwidth}
  \strut\newline
  \centering
    \includegraphics[width=0.8\textwidth]{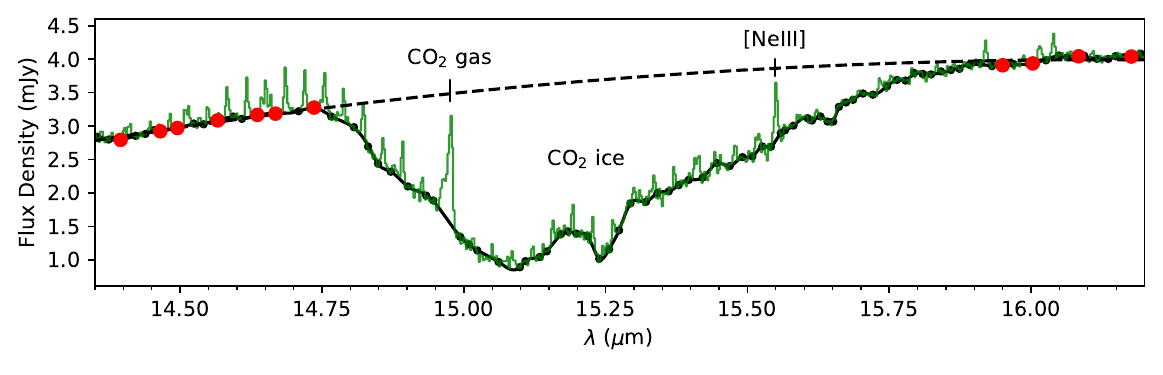}
    \includegraphics[width=0.8\textwidth]{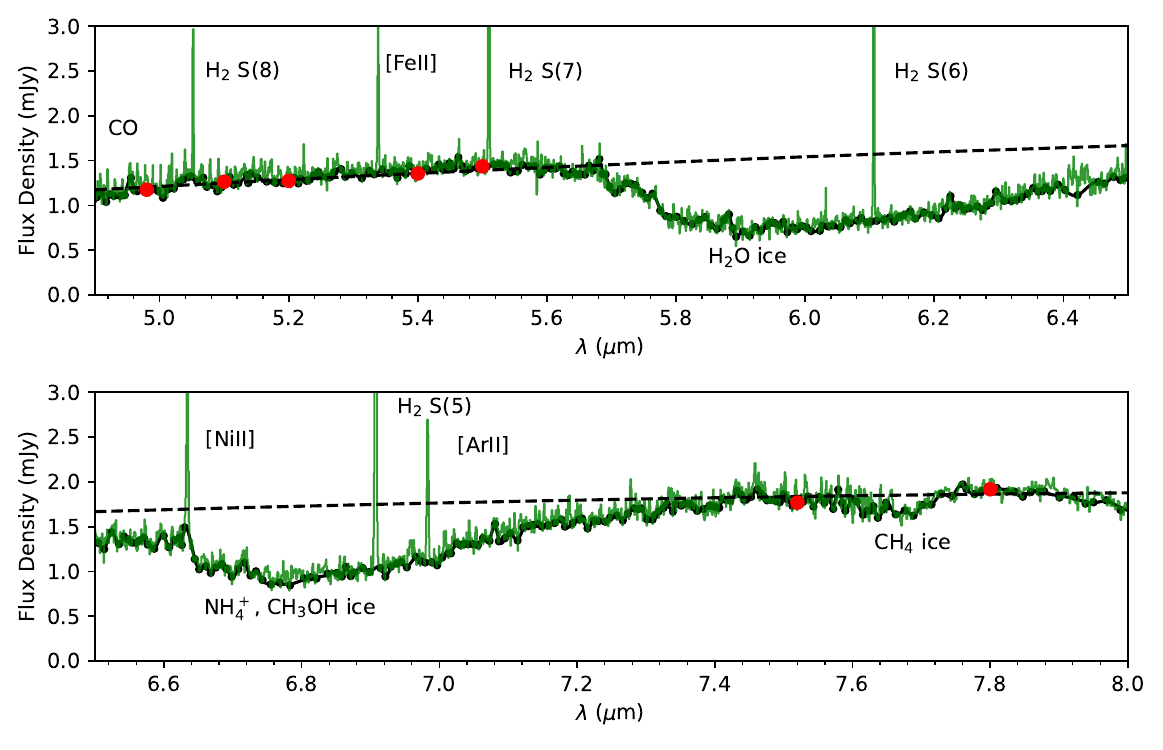} 
  \captionof{figure}{Example of local continuum spline fitting and modelling of the CO$_2$ (top panel), H$_2$O (middle panel), and NH$_4^{+}$ (bottom panel) ice absorption feature's optical depth. The dashed line shows the model for the thermal continuum fit to red points on either side of the ice feature, while the solid black line indicates the cubic spline fit. The data is shown in green for source A. The majority of unlabelled emission features in the lower panel are H$_2$O emission.}\label{fig:cont_spline}
\end{minipage}

\FloatBarrier
\section{Model fits}
\label{sec:app_model_fits}

\begin{figure*}
    \centering
    \includegraphics[width=0.9\textwidth]{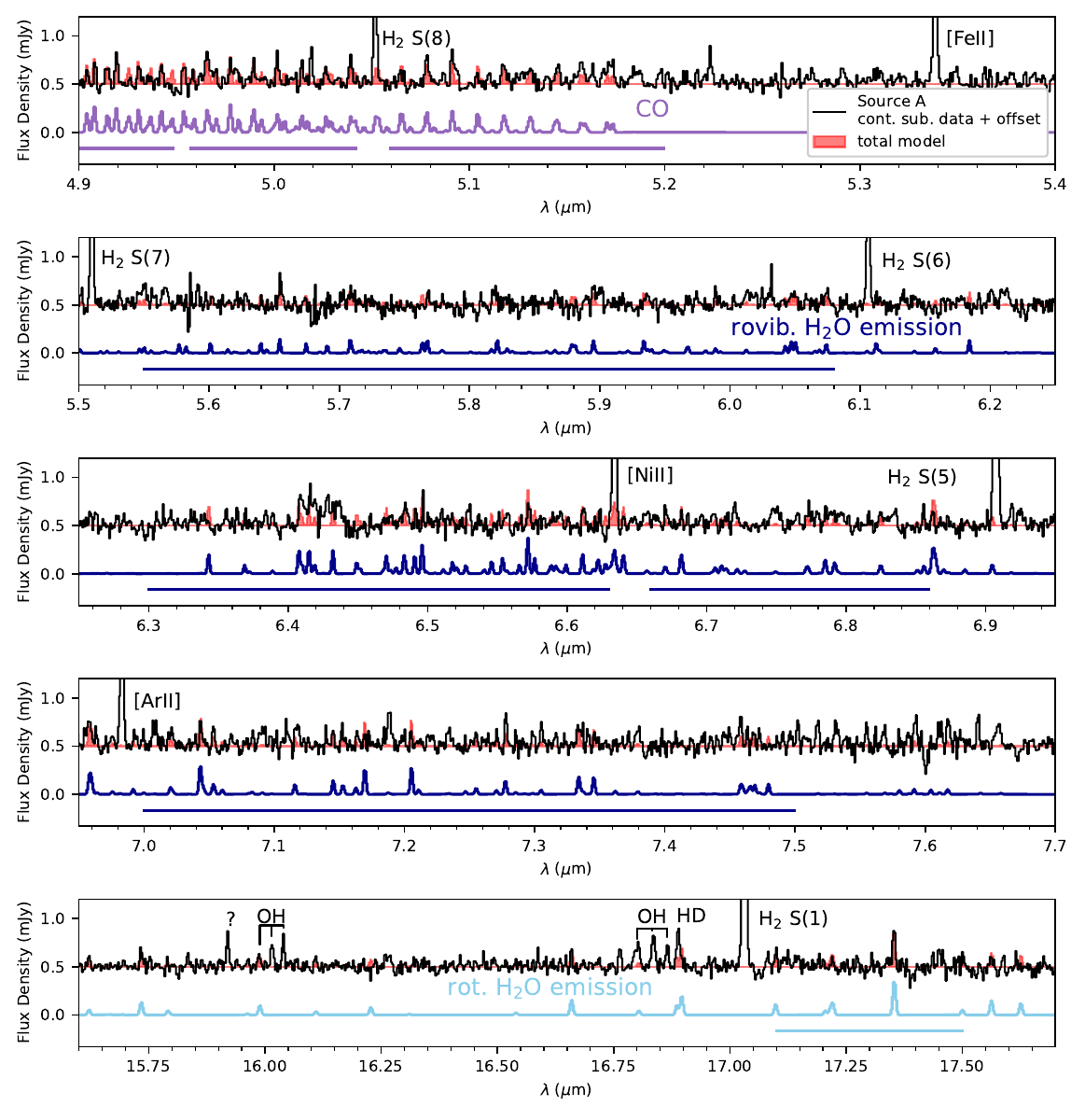}
    \caption{As Figure \ref{fig:total_model_spectra_AB}, but for the best-fit LTE models of CO, H$_2$O emission at 5.5-7.7 $\mu$m and 16-18 $\mu$m in source A.}
    \label{fig:total_model_spectra_A_H2O}
\end{figure*}

\begin{figure*}
    \centering
    \includegraphics[width=0.9\textwidth]{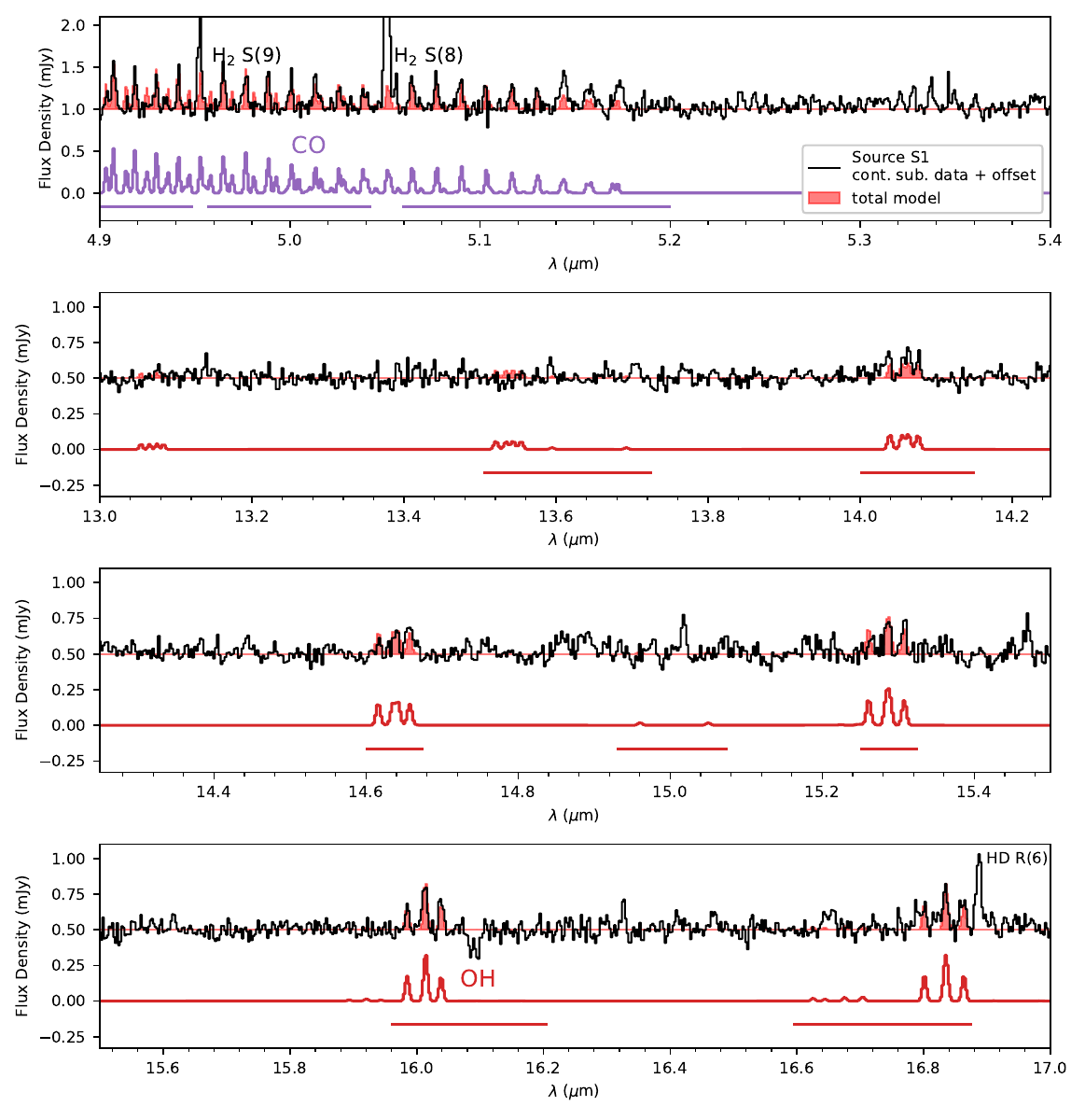}
    \caption{As Figure \ref{fig:total_model_spectra_AB}, but for the best-fit LTE model of CO and OH emission in source S1.}
    \label{fig:total_model_spectra_N}
\end{figure*}

\begin{figure*}
    \centering
    \includegraphics[width=0.9\textwidth]{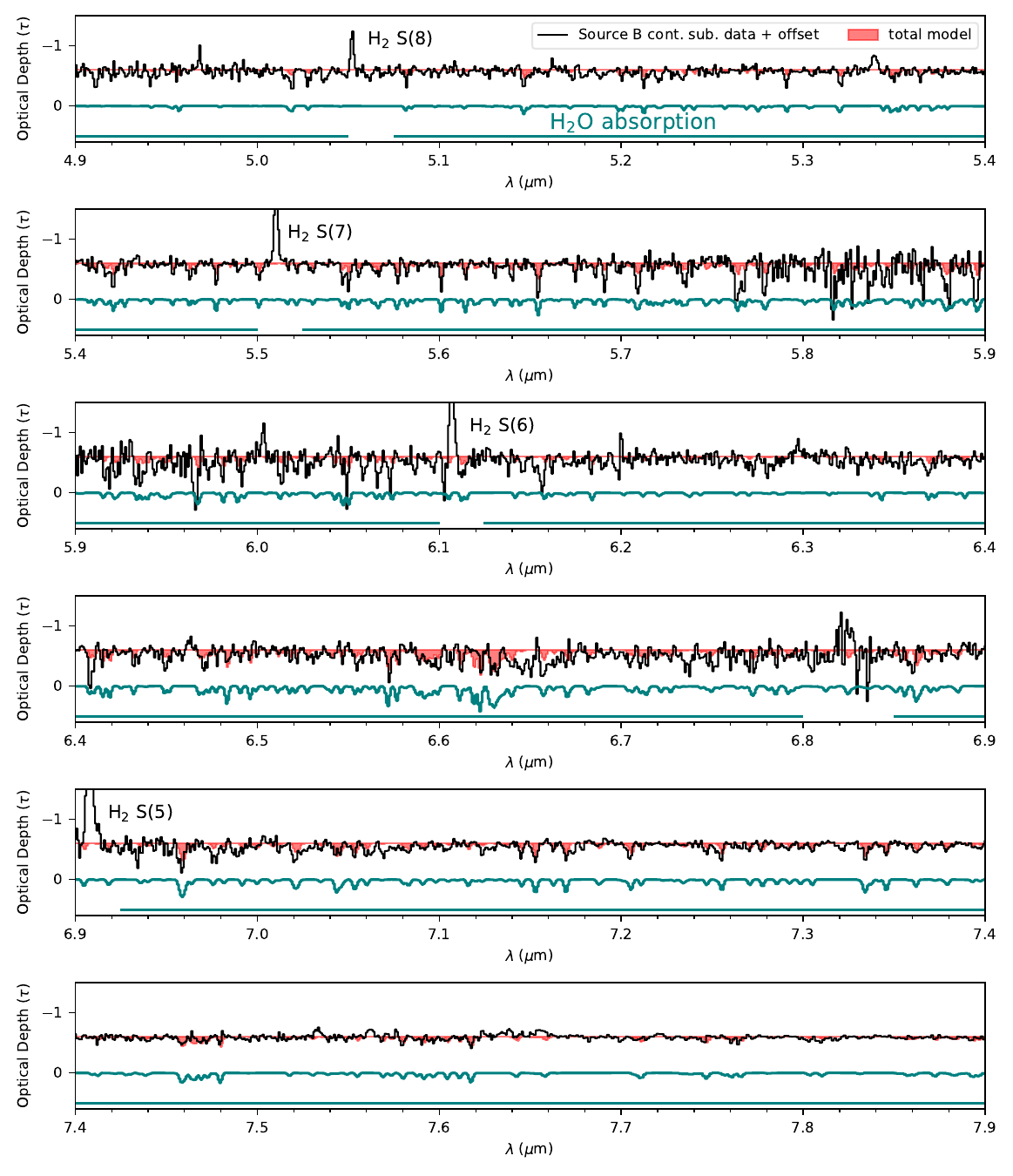}
    \caption{As Figure \ref{fig:total_model_spectra_AB}, but the observed data for source B is shown in units of optical depth with an offset of -0.5 relative to the best-fit LTE model for H$_2$O absorption.}
    \label{fig:total_model_spectra_B_H2O}
\end{figure*}

In this section, we show the best-fit LTE models for each molecule in the same manner as Figure \ref{fig:total_model_spectra_AB}. In Figures, \ref{fig:total_model_spectra_A_H2O}, \ref{fig:total_model_spectra_N}, and \ref{fig:total_model_spectra_B_H2O},  the LTE model for each fit is shown by the coloured lines, and compared to the continuum-subtracted data (black) in the overlaid red-shaded region.

\FloatBarrier
\section{LTE model $\chi^2$ maps}
\label{sec:app_chi2_maps}

\begin{figure*}[htb]
    \centering
    \includegraphics[scale=0.65]{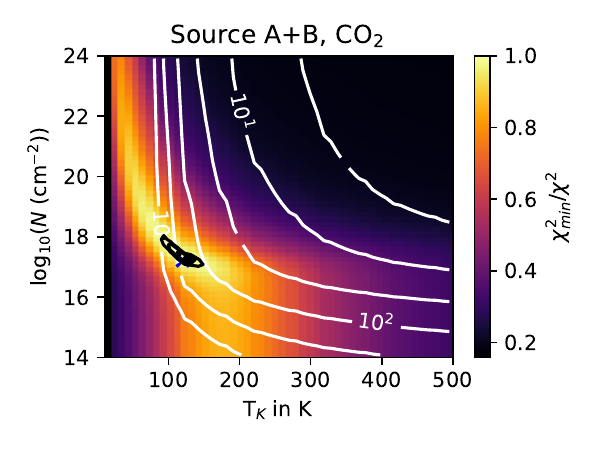}
    \includegraphics[scale=0.65]{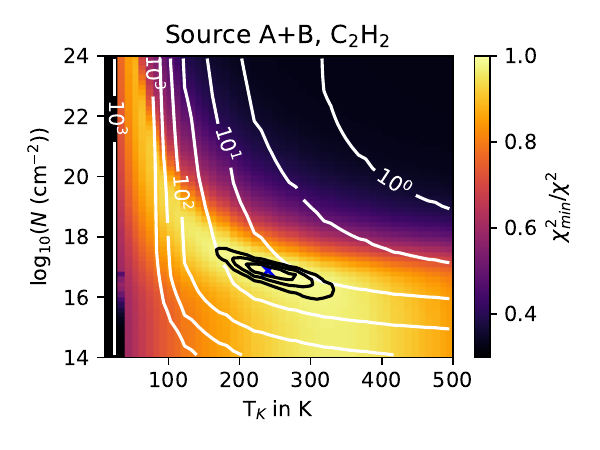}
    \includegraphics[scale=0.65]{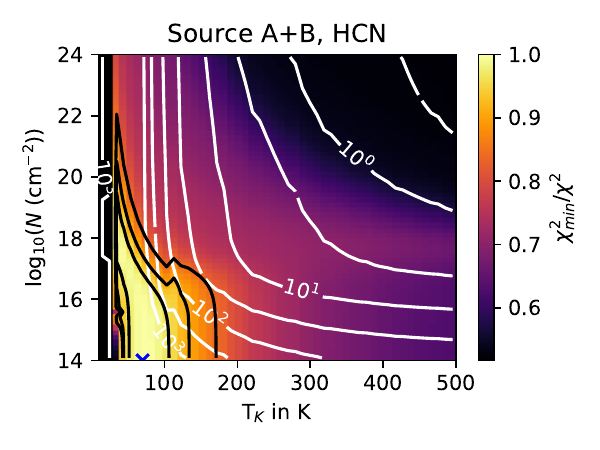}
    \includegraphics[scale=0.65]{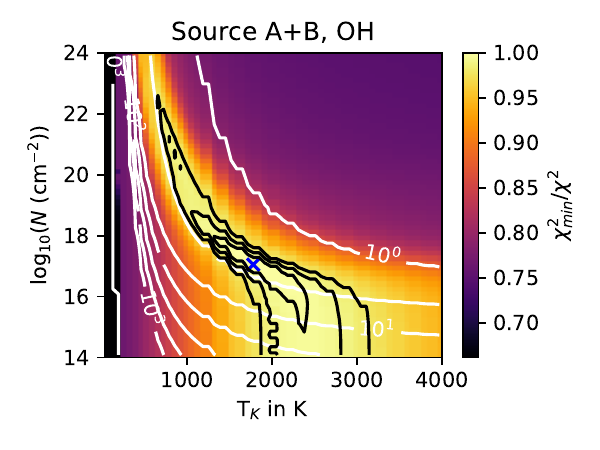}
    
    \caption{Normalized $\chi^2$ maps (colour scale) of the LTE model temperature and column density grids for CO$_2$, C$_2$H$_2$, HCN, 
    and OH model fits in source A+B. The best-fit emitting radius for each model is overlaid as white contours, while the 1, 2, and 3 $\sigma$ confidence interval contours are overlaid in black. The best-fit model at the minimum $\chi^2$ is shown by a blue cross.}
    \label{fig:chi2maps_sourceAB}
\end{figure*}

\begin{figure*}[htb]
    \centering
    \includegraphics[scale=0.65]{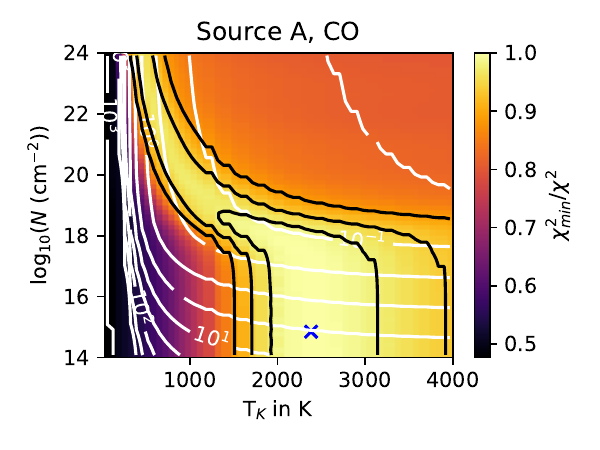}
    \includegraphics[scale=0.65]{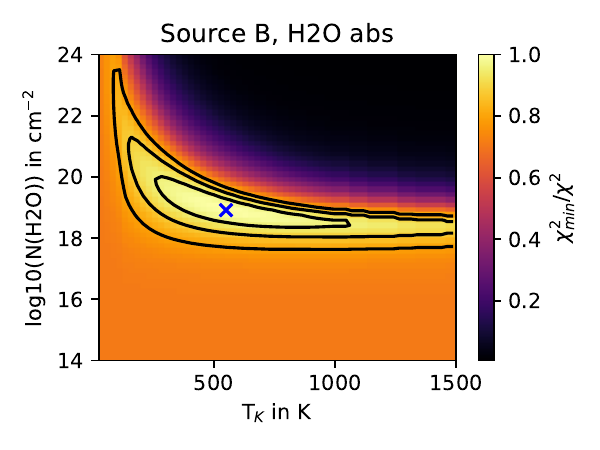}
    \caption{As Figure \ref{fig:chi2maps_sourceAB}, but for CO emission in source A  and H$_2$O absorption in source B}.
    \label{fig:chi2maps_sourceAandB}
\end{figure*}

\begin{figure*}[htb]
    \centering
    \includegraphics[scale=0.65]{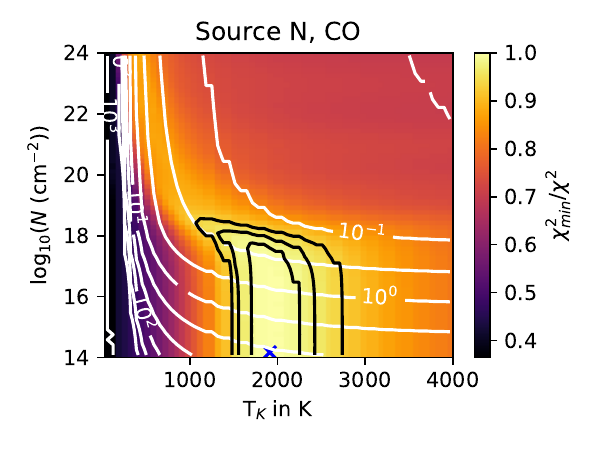}
    \includegraphics[scale=0.65]{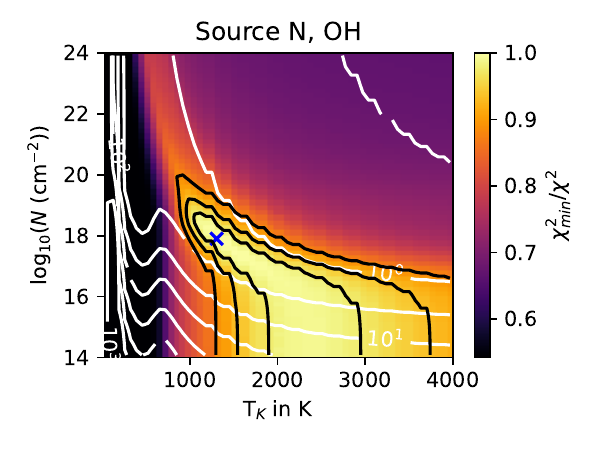}    
    \caption{As Figure \ref{fig:chi2maps_sourceAB}, but for CO and OH emission in source S1.}
    \label{fig:chi2maps_sourceN}
\end{figure*}

The resolution of the LTE model grids used to fit our observations are the same in column density, but vary in temperature depending on the $S/N$ of the emission and excitation conditions. For all
species, the column density range is $N=10^{14} - 10^{24}$ cm$^{-2}$ in steps of $\log(N)=0.166$. For the relatively high $S/N$ emission from cold CO$_2$, HCN, and C$_2$H$_2$, the temperature range is $T=10 - 500$ K in steps of 10 K. For the highly excited emission from CO and OH, a range of $T=10 - 500$ K in steps of 67 K is used. For all other species, the the temperature in the grid ranges from $T=25 - 1500$ in steps of 25K. Our LTE model fits are shown for source A+B in Figure \ref{fig:chi2maps_sourceAB}, for source A and B separately in Figure \ref{fig:chi2maps_sourceAandB}, and for source S1 in Figure \ref{fig:chi2maps_sourceN}.

\end{appendix}
\end{document}